\documentclass[
 reprint,
 amsmath,amssymb,
 aps,pra,superscriptaddress
]{revtex4-2}

\usepackage{graphicx}
\usepackage{dcolumn}
\usepackage{bm}
\usepackage{color}
\usepackage{empheq}

\usepackage{hyperref}
\hypersetup{
colorlinks=false,
hidelinks
}

\let\p\partial

\let\f\frac
\def\be{\begin{equation}}\def\ee{\end{equation}}
\def\bea{\begin{eqnarray}}\def\eea{\end{eqnarray}}

\graphicspath{{Figures/}}

\begin{document}

\preprint{APS/123-QED}

\title{Corotation of two quantized vortices coupled with collective modes in self-gravitating Bose-Einstein condensates}

\author{Kenta Asakawa}
\email{sn22896p@st.omu.ac.jp}
\affiliation{Department of Physics, Osaka Metropolitan University, 3-3-138 Sugimoto, Sumiyoshi-Ku, Osaka 558-8585, Japan}
\author{Makoto Tsubota}
\email{tsubota@omu.ac.jp}
\affiliation{Department of Physics, Osaka Metropolitan University, 3-3-138 Sugimoto, Sumiyoshi-Ku, Osaka 558-8585, Japan}
\affiliation{Nambu Yoichiro Institute of Theoretical and Experimental Physics (NITEP), Osaka Metropolitan University, 3-3-138 Sugimoto, Sumiyoshi-Ku, Osaka 558-8585, Japan}

\date{\today}

\begin{abstract}
We numerically examine the corotation of two parallel quantized vortices in a self-gravitating Bose-Einstein condensate (BEC) employing the Gross-Pitaevskii-Poisson equations. 
The long-range gravitationally attractive interaction allows the BEC to self-confine without the need for external potentials, while the density-dependence of the gravitational potential induces intriguing behaviors in the quantized vortices. 
The aim of this study is to provide a clue for understanding the corotation of two quantized vortices under the influence of gravitational interactions.
The corotation of two quantized vortices is coupled with collective modes of the BEC, which markedly differs from the behavior observed in typical BECs confined by an external potential.
The rotational period increases linearly with the initial position from the center of the BEC.
This deviation from the quadratic increase observed in a uniform BEC suggests that the gravitational interaction exerts a drag effect on the rotating quantized vortices. 
The two closely positioned quantized vortices rotate along elliptical orbits with radial fluctuations.
However, when the quantized vortices are initially positioned beyond a critical radius comparable to their core sizes, their trajectory transitions into an outward spiral, implying the onset of effective dissipation.
Our findings demonstrate that the radial fluctuations of the quantized vortex resonate with the quadrupole mode of the BEC, giving rise to a dissipation mechanism.
\end{abstract}

\maketitle

\section{\label{sec:level1}INTRODUCTION}
A quantized vortex, characterized by quantized circulation, plays a crucial role in the distinctive dynamics of Bose-Einstein condensates (BECs), superfluid $^4$He, $^3$He, and other quantum fluids \cite{fetter2009}.
It appears as a topological defect of the macroscopic wavefunction that describes the ordered state of the quantum fluid.
Extensive theoretical and experimental research on the dynamics of the quantized vortex has been conducted in low-temperature physics, contributing significantly to the current understanding of some characteristic phenomena, such as the vortex lattice formation \cite{ketterle2001,kasamatsu2002} and quantum turbulence \cite{tsatsos2016, tsubota2013}.
Particularly, the behavior of the quantized vortices in the atomic BEC has been theoretically explored using the Gross-Pitaevskii (GP) model based on the mean-field approximation \cite{Gross1963, Pitaevskii1961}.

Inter-particle interactions are intricately connected to the behavior of the BEC and the motions of quantized vortices; the properties of the quantum fluid are significantly influenced by variations in the strength, range, and anisotropy of these interactions.
In a typical atomic BEC, particles repel each other in short ranges due to contact interactions.
This repulsive short-range interaction can be modulated through the Feshbach resonance to become attractive \cite{Inouye1998}, potentially leading to instabilities in the BEC and the quantized vortices within it \cite{PethickSmith, Lundh2004}.
Furthermore, in a dipolar BEC, dipole-dipole interactions, which vary in magnitude and direction depending on their orientation between particles, act over a wide range \cite{Lahaye2009, Chomaz2023}.
The properties such as the long-range interactions and the switching between attraction and repulsion depending on the orientation between particles induce transverse instabilities in the BEC and straight quantized vortex \cite{Klawunn2008, Cidrim2018}, as well as the shear motion in the Tkachenko mode exhibited by the striped vortex lattice \cite{Jia2018}.
The latest study investigates the impact of the dipole-dipole interactions on the generation of quantized vortices and the development of quantum turbulence by stirring the quasi-two-dimensional dipolar BEC with a rotating obstacle \cite{sabari2024}.

While low-temperature physics focuses on the atomic BEC where the contact and dipole-dipole interactions operate between atoms, astrophysics has paid attention to gravitationally interacting BECs composed of the ultralight scalar particle \cite{Hu2000, matos2001, bohmer2007, chavanis2011_1, niemeyer2020, Hui2021, ferreira2021, Matos2024}.
This is because such particles or BECs can be a plausible candidate for dark matter which constitutes the unidentified and primary component of the matter in the present universe.
A BEC affected by the gravitational interaction is defined as a self-gravitating BEC. 
It is described by the non-local GP model coupled with the Newtonian gravitational field generated by the BEC itself.
Since the gravitational interaction acts isotropically as an attractive long-range interaction, a self-gravitating BEC is confined by its intrinsic gravitational potential and can attain equilibrium independently, without the necessity of an external potential, distinguishing it from conventional atomic BECs \cite{ODell2000, guzman2006}.

Notably, several optical methods have been proposed to simulate the quantum fluid properties of such a self-gravitating BEC in laboratory systems.
These include inducing the "gravitation-like" interatomic interaction using off-resonant laser \cite{ODell2000}, and analogies with laser light propagating through thermo-optical media \cite{Paredes2020}.

The quantum-fluid dynamics of self-gravitating BECs with quantized vortices are intriguing from both the perspectives of low-temperature physics and astrophysics. 

From the perspective of low-temperature physics, there is significant interest in the alterations in the behavior exhibited by BECs and their quantized vortices under the influence of isotropic long-range gravitational interactions.
Their dynamics become nontrivial due to the density dependence of its own gravitational potential. 
For instance, the oscillation frequency of the collective modes in a self-gravitating BEC in response to small fluctuations depends on the total mass of the BEC \cite{giovanazzi2001, asakawa2024}, reflecting the density dependence of the gravitational potential; this is different from an atomic BEC trapped by external potentials \cite{stringari, PethickSmith}. 
When quantized vortices are present in a self-gravitating BEC, the change in the density profile caused by the vortex core deforms the gravitational potential \cite{nikolaieva2023}, and both the BEC and quantized vortices move within this deformed potential. 
The intricate nonlinearity characteristic of self-gravitating BECs is likewise observed in dipolar BECs with long-range interactions, despite the differences in the anisotropy of these interactions.

On the other hand, quantum fluid phenomena driven by quantized vortices in a self-gravitating BEC could serve as a touchstone for confirming the possibility that the BEC of ultralight scalar particles constitutes dark matter, as inferred from observations of galaxies and galaxy clusters. 
Astrophysical interest in this issue has been longstanding, encompassing concepts such as the possibility of the formation of vortex lattices in the ultralight dark matter BEC \cite{zinner2011}, comparisons between the critical angular velocity of a BEC with a central vortex and the rotation speed of galaxies \cite{kain2010}, and the deformation of configurations due to quantized vortices \cite{daller2012}. 
Recently, three-dimensional numerical studies have explored the stability of the self-gravitating BEC with a single central vortex \cite{nikolaieva2021, dmitriev2021}, the gravitational wave emission during head-on collisions of two self-gravitating BECs with quantized vortices \cite{nikolaieva2023}, and quantum turbulence resulting from the merger of multiple self-gravitating BECs \cite{mocz2018, liu2023, barenghi2023}.

As an initial step toward understanding the dynamics of quantized vortices influenced by gravitational interactions in BECs, this study undertakes a numerical investigation of the orbital rotation of two quantized like-charged vortices in a three-dimensional self-gravitating BEC. 
There are numerous studies on the behavior of a few vortices in quantum fluids \cite{fetter2009, navarro2013, gautam2014, nakamura2016}, and it is well-known that two parallel, straight vortices corotate within the quantum fluid, following the flow fields generated by each other. 
In a uniform quantum fluid at zero temperature, the two quantized vortices rotate while maintaining their initial distance, and the rotational period increases proportionally to the square of the distance. 
On the other hand, when dissipation is present to reduce the rotational energy, the two parallel vortices spiral out as they rotate. 
Their corotational behavior is expected to be similarly exhibited even in a self-gravitating BEC, although the gravitational interaction, or the gravitational potential, is anticipated to influence their rotational speed and trajectory. 
Actually, recent studies on the rotation of two quantized vortices in a dipolar BEC, which also exhibits the long-range interaction, have revealed the non-circular rotational trajectories and the dependence of the rotational speed on the dipole-dipole interaction \cite{zhao2021, zhao2022}.

We compose this paper as follows.
First, the formalism of the self-gravitating BEC and the numerical method based on it are shown in Sec.$\mathrm{I}\hspace{-1.2pt}\mathrm{I}$.
Our numerical results are shown in Sec.$\mathrm{I}\hspace{-1.2pt}\mathrm{I}\hspace{-1.2pt}\mathrm{I}$.
Finally, we conclude this paper in Sec.$\mathrm{I}\hspace{-1.2pt}\mathrm{V}$.

\section{THEORETICAL MODEL}
\subsection{FORMULATION OF THE SELF-GRAVITATING BEC AND THE QUANTIZED VORTEX}

When the self-gravitating BEC is composed of the Bose particles of the mass $m$ and the s-wave scattering length $a$ near zero temperature, its time evolution is governed by the Gross-Pitaevskii-Poisson (GPP) equations defined as
\be
  \begin{split}
     i\hbar\f{\p {\psi(\bm{r},t)}}{\p t}
      =& -\f{\hbar^2}{2m}\nabla^2\psi(\bm{r},t)+mV(\bm{r},t){\psi(\bm{r},t)} \\
      & +\f{4\pi\hbar^2a}{m} |\psi(\bm{r},t)|^2{\psi(\bm{r},t)},
  \end{split}
  \label{gp}
\ee
and
\be
{\nabla}^2V(\bm{r},t)= 4\pi Gm\lvert{\psi(\bm{r},t)}\rvert^2,
\label{poisson}
\ee
where, $\psi(\bm{r},t)$ denotes the macroscopic wavefunction and $V(\bm{r},t)$ represents the gravitational potential \cite{bohmer2007, chavanis2011_1, ferreira2021}.
The macroscopic wavefunction can be rewritten using the Madelung representation $\psi(\bm{r},t)=\sqrt{\rho(\bm{r},t)/m}\exp[i\theta(\bm{r},t)]$, where $\rho(\bm{r},t)$ denotes the mass density and $\theta(\bm{r},t)$ is the phase. 
The velocity field is introduced as $\bm{v}(\bm{r},t)=\hbar\nabla\theta(\bm{r},t)/m$.
Eq. (\ref{poisson}) is the Poisson equation and determines the gravitational potential in Eq. (\ref{gp}) from the density profile of the BEC.
Hence, the GPP equations exhibit that the self-gravitating BEC is constrained by and evolves according to its gravitational potential which depends on the density profile of the BEC itself.

Although the velocity field of the BEC is ordinarily irrotational, the quantized vortex defined as the topological deficit locally gives rise to the rotational flow in the BEC.
For example, when the phase is given by $\theta(\bm{r},t)=\tan^{-1}(y/x)$ using the Cartesian coordinates $(x,y,z)$, the counter-clockwise velocity field is generated around the z-axis and its amplitude $|\bm{v}|\propto1/\sqrt{x^2+y^2}$, which displays a straight vortex line along the $z$-axis.
The quantized vortex is characterized by the discretized (quantized) circulation in units of $\kappa=2\pi\hbar/m$ and the core size comparable to the coherence length $\xi=\sqrt{m/(8\pi\rho_ca)}$, where $\rho_c$ denotes a characteristic value of the mass density.
Note that the quantized vortex whose circulation exceeds the unit hardly appears because it is energetically unstable even in the self-gravitating BEC \cite{nikolaieva2021, dmitriev2021}.

The total energy $E(t)$ of the GPP equations (\ref{gp}) and (\ref{poisson}) is given by the sum of the kinetic energy $E_\mathrm{k}(t)$, gravitational energy $E_\mathrm{p}(t)$, and contact interaction energy $E_\mathrm{i}(t)$, namely $E(t)=E_\mathrm{k}(t)+E_\mathrm{p}(t)+E_\mathrm{i}(t)$.
These energy components are written by
\be
E_\mathrm{k}(t)=\f{\hbar^2}{2m}\int d\bm{r}\left|\nabla\psi(\bm{r},t)\right|^2,
\label{kinetic_energy}
\ee
\be
E_\mathrm{p}(t)=\f{m}{2}\int d\bm{r}V(\bm{r},t)\left|\psi(\bm{r},t)\right|^2,
\label{gravitational_energy}
\ee
and
\be
E_\mathrm{i}(t)=\f{2\pi\hbar^2a}{m}\int d\bm{r}\left|\psi(\bm{r},t)\right|^4.
\label{contact_energy}
\ee
Using the Madelung representation, the kinetic energy is divided into two components: the quantum pressure component $E_\mathrm{q}(t)$ and the velocity component $E_\mathrm{v}(t)$ given by
\be
E_\mathrm{q}(t)=\f{\hbar^2}{2m^2}\int d\bm{r}\left|\nabla\left(\sqrt{\rho(\bm{r},t)}\right)\right|^2,
\label{quantum_energy}
\ee
and
\be
E_\mathrm{v}(t)=\f{1}{2}\int d\bm{r}\rho(\bm{r},t)\bm{v}(\bm{r},t)^2.
\label{velocity_energy}
\ee
$E_\mathrm{v}$ is, furthermore, divided into the compressible part 
\be
E_\mathrm{vc}(t)=\f{1}{2}\int d\bm{r}\rho(\bm{r},t)\bm{v}_\mathrm{c}(\bm{r},t)^2,
\label{compressible_energy}
\ee
 and incompressible part
 \be
E_\mathrm{vi}(t)=\f{1}{2}\int d\bm{r}\rho(\bm{r},t)\bm{v}_\mathrm{i}(\bm{r},t)^2,
\label{incompressible_energy}
\ee
based on the Helmholtz decomposition \cite{nore1997_1, nore1997_2}.
Here, the compressible velocity field $\bm{v}_\mathrm{c}(\bm{r},t)$ and the incompressible one $\bm{v}_\mathrm{i}(\bm{r},t)$ are respetively satisfied that $\nabla\times\bm{v}_\mathrm{c}(\bm{r},t)=\bm{0}$ and $\nabla\cdot\bm{v}_\mathrm{i}(\bm{r},t)=0$, and $\bm{v}_\mathrm{c}(\bm{r},t)+\bm{v}_\mathrm{i}(\bm{r},t)=\bm{v}(\bm{r},t)$.
Although the deformation/oscillation of the BEC can contribute to both, the quantized vortices contribute only to the incompressible part.

\subsection{NUMERICAL METHOD}
To carry out the numerical study of the self-gravitating BEC, we solve the dimensionless GPP equations,
\be
\begin{split}
	i\f{\partial\tilde{\psi}(\tilde{\bm{r}},\tilde{t})}{\partial\tilde{t}}
	=& -{\frac{1}{2}\tilde{\nabla}^2\tilde{\psi}(\tilde{\bm{r}},\tilde{t})+\tilde{V}(\tilde{\bm{r}},\tilde{t})\tilde{\psi}(\tilde{\bm{r}},\tilde{t})} \\
	& +\tilde{a}\lvert{\tilde{\psi}(\tilde{\bm{r}},\tilde{t})}\rvert^2\tilde{\psi}(\tilde{\bm{r}},\tilde{t}),
\end{split}
\label{dimensionless_gp}
\ee
and
\be
\tilde{\nabla}^2\tilde{V}(\tilde{\bm{r}},\tilde{t}) = \lvert{\tilde{\psi}(\tilde{\bm{r}},\tilde{t})\rvert}^2
\label{dimensionless_poisson}
\ee
in a three-dimensional numerical box.
Here, the dimensionless rules are respectively $\tilde{\bm{r}}=\tilde{\lambda}(mc/\hbar)\bm{r}$, $\tilde{t}=\tilde{\lambda}^2(mc^2/\hbar)t$, $\tilde{\psi}=\{\sqrt{4\pi G}\hbar/(\sqrt{m}c^2\tilde{\lambda}^2)\}\psi$, $\tilde{V}=V/(\tilde{\lambda}^2c^2)$, and $\tilde{a}=\{\tilde{\lambda}^2c^2/(mG)\}a$, where $c$ is the velocity of light in the vacuum \cite{guzman2014, asakawa2024}. 
The criterion of these rules is the Compton wavelength, and the scaling parameter $\tilde{\lambda}$ makes such dimensionless variables suitable without changing the formulations of Eqs. (\ref{dimensionless_gp}) and (\ref{dimensionless_poisson}).

The dimensionless GP equation of Eq. (\ref{dimensionless_gp}) is solved by the pseudo-spectrum method and the fourth-order Runge-Kutta method.
Using the pseudo-spectrum method, the periodic boundary conditions must be applied to Eq. (\ref{dimensionless_gp}).
On the other hand, the dimensionless Poisson equation of Eq. (\ref{dimensionless_poisson}) can be solved by the Poisson solver in the Intel MKL.
Considering the anisotropic deformation of the BEC near the center of the numerical box, we can employ the multipole expansion solution of the gravitational potential until the second-order terms, written by 
\be
\begin{split}
\tilde{V}(\tilde{\bm{r}},\tilde{t})
=&
-\f{\tilde{M}}{4\pi\tilde{r}}
-\f{1}{4\pi\tilde{r}^3}\sum_{i\in\{x,y,z\}}\tilde{P}_i\tilde{r}_i \\
&-\f{3}{8\pi\tilde{r}^5}\sum_{i,j\in\{x,y,z\}}\tilde{I}_{ij}\tilde{r}_i\tilde{r}_j,
\label{dimensionless_multipole}
\end{split}
\ee
as the boundary conditions for Eq. (\ref{dimensionless_poisson}), where $\tilde{r}_i$ is $i$-component of $\tilde{\bm{r}}$, \textit{e.g.} $\tilde{r}_x=\tilde{x}$, and $\tilde{r}=|\tilde{\bm{r}}|$.
$\tilde{M}=\int d\tilde{\bm{r}}|\tilde{\psi}(\tilde{\bm{r}},\tilde{t})|^2$ represents the dimensionless total mass, $\tilde{P}_i=\int d\tilde{\bm{r}}\tilde{r}_i|\tilde{\psi}(\tilde{\bm{r}},\tilde{t})|^2$ represents the $i$-component of the dimensionless dipole moment, and $\tilde{I}_{i,j}=\int d\tilde{\bm{r}}(\tilde{r}_i\tilde{r}_j-\delta_{i,j}\tilde{\bm{r}}^2/3) |\tilde{\psi}(\tilde{\bm{r}},\tilde{t})|^2$ represents $(i,j)$-component of the dimensionless quadrupole moment.

The imaginary-time evolution of the GPP equations is applied to prepare the initial state.
In this scheme, we initially set a suitable configuration and renew it by solving
\be
\begin{split}
	-\f{\partial\tilde{\psi}(\tilde{\bm{r}},\tilde{t})}{\partial\tilde{t}}
	=& -{\frac{1}{2}\tilde{\nabla}^2\tilde{\psi}(\tilde{\bm{r}},\tilde{t})+\tilde{V}(\tilde{\bm{r}},\tilde{t})\tilde{\psi}(\tilde{\bm{r}},\tilde{t})} \\
	& +\tilde{a}\lvert{\tilde{\psi}(\tilde{\bm{r}},\tilde{t})}\rvert^2\tilde{\psi}(\tilde{\bm{r}},\tilde{t})-\tilde{\mu}(\tilde{t})\tilde{\psi}(\tilde{\bm{r}},\tilde{t}),
\end{split}
\label{dimensionless_igp}
\ee
and the dimensionless Poisson equation of Eq. (\ref{dimensionless_poisson}) until the system energically relaxes enough.
The diffusion equation of Eq. (\ref{dimensionless_igp}) is obtained by a replacement $\tilde{t}\rightarrow -i\tilde{t}$ of Eq. (\ref{dimensionless_gp}) and the chemical potential $\tilde{\mu}(\tilde{t})$ allows the system to relax while conserving the total mass.
In the imaginary-time evolution, we employ the pseudo-spectrum method and Nesterov's accelerated gradient method \cite{nesterov1983, donoghue2015, su2016} to implement that more rapidly.

We prepare the initial state by positioning two parallel quantized vortices along the z-axis, aligned on the x-axis.
The starting configuration of the imaginary-time evolution is written by $\tilde{\psi}(\tilde{\bm{r}},0)=\tilde{f}(\tilde{\bm{r}})\exp[i\theta(\tilde{\bm{r}})]$.
The function $\tilde{f}(\tilde{\bm{r}})$ is chosen such that it satisfies $\tilde{f}(\tilde{x}_{v,1},0,\tilde{z})=\tilde{f}(\tilde{x}_{v,2},0,\tilde{z})=0$, where $\tilde{x}_{v,1}$ and $\tilde{x}_{v,2}$ are the $x$-coordinates of the quantized vortices, and the phase is set as
\be
\theta (\tilde{\bm{r}})
=
\tan^{-1}\left(\f{\tilde{y}}{\tilde{x}-\tilde{x}_{v,1}}\right)
+
\tan^{-1}\left(\f{\tilde{y}}{\tilde{x}-\tilde{x}_{v,2}}\right).
\label{initial_phase}
\ee

Two conditions are imposed during the imaginary-time evolution of the GPP equations.
First, the gravitational potential is symmetrized in $xy$, $yz$, and $zx$ planes, namely $\tilde{V}(\tilde{x},\tilde{y},\tilde{z},\tilde{t})=\tilde{V}(\tilde{x},\tilde{y},-\tilde{z},\tilde{t})$, $\tilde{V}(\tilde{x},\tilde{y},\tilde{z},\tilde{t})=\tilde{V}(-\tilde{x},\tilde{y},\tilde{z},\tilde{t})$, and $\tilde{V}(\tilde{x},\tilde{y},\tilde{z},\tilde{t})=\tilde{V}(\tilde{x},-\tilde{y},\tilde{z},\tilde{t})$, in order to fix the center of mass of the BEC.
Thus, we consider that the quantized vortices are symmetrically located in the $yz$ plane, $\tilde{x}_{v,1}=-\tilde{x}_{v,2}=\tilde{r}_o$, where $\tilde{r}_o$ is the initial distance from the center of the BEC.
Second, two initial identities, 
\be
\mathrm{Re}\left[\tilde{\psi}(\tilde{\bm{r}},\tilde{t})\right]=0 ~~~~\mathrm{when}~\theta (\tilde{\bm{r}})=0,~\pm\pi, 
\label{fix_vortex_re}
\ee
and 
\be
\mathrm{Im}\left[\tilde{\psi}(\tilde{\bm{r}},\tilde{t})\right]=0 ~~~~\mathrm{when}~\theta (\tilde{\bm{r}})=\pm\f{\pi}{2}
\label{fix_vortex_im}
\ee
are kept in order to fix the positions of the quantized vortices.
These are the expansion of the conditions used in \cite{abad2009} which considers the case of one straight vortex.
Adapting Eq. (\ref{initial_phase}) and $\tilde{x}_{v,1}=-\tilde{x}_{v,2}=\tilde{r}_o$ to Eqs. (\ref{fix_vortex_re}) and (\ref{fix_vortex_im}), these conditions in our situation are particularly rewritten as $\mathrm{Re}[\tilde{\psi}(\tilde{\bm{r}},\tilde{t})]=0$ on a hyperbola $\tilde{x}^2-\tilde{y}^2=\tilde{r}_o^2$ and $\mathrm{Im}[\tilde{\psi}(\tilde{x},0,\tilde{z},\tilde{t})]=0$.

Finally, we show the physical and numerical parameters for this work.
We consider that the boson has the mass $m=3\times 10^{-24}~\mathrm{eV}$ and the s-wave scattering length $a\approx3.69\times 10^{-81}~\mathrm{m}$ referred in \cite{nikolaieva2021, asakawa2024}, and the total mass of the BEC is $3\times 10^{12}M_\odot$ which is comparable to the typical mass of the galaxy \cite{Binney}, where $M_\odot$ denotes the solar mass.
The numerical box has the length $40$ and the number of glids $256^3$, namely the spatial resolution is $d\tilde{x}=40/256=0.15625$ or $dx\approx0.10~\mathrm{kpc}$ based on the dimensionless rule.
The time resolution is $10^{-3}$.
We determine the scaling parameter $\tilde{\lambda}$ such that the dimensionless Thomas-Fermi radius $\tilde{R}_\mathrm{TF}$ is $10$.
The Thomas-Fermi radius characterizes the size of the self-gravitating BEC and is written as $R_\mathrm{TF}=\pi\sqrt{\hbar^2a/(Gm^3)}$ \cite{bohmer2007}.
As a result, $\tilde{\lambda}=10/(\pi c)\sqrt{Gm/a}\approx 3.30\times 10^{-3}$.
The initial distance of the quantized vortex from the center of the BEC is set as $r_o/dx=5,~8,~10,~12,~13,~15,~20,~25,~30,~40$, namely $r_o\approx0.5,~0.8,~1.0,~1.2,~1.3,~1.5,~2.0,~2.5,~3.0,~4.0$ in a unit of kilopersec.
Note that all of $r_o$ is smaller than $R_\mathrm{TF}\approx64dx$, which means that the vortices are put inner the BEC.

\section{RESULTS}
\begin{figure*}
\centering
\includegraphics[width=18cm]{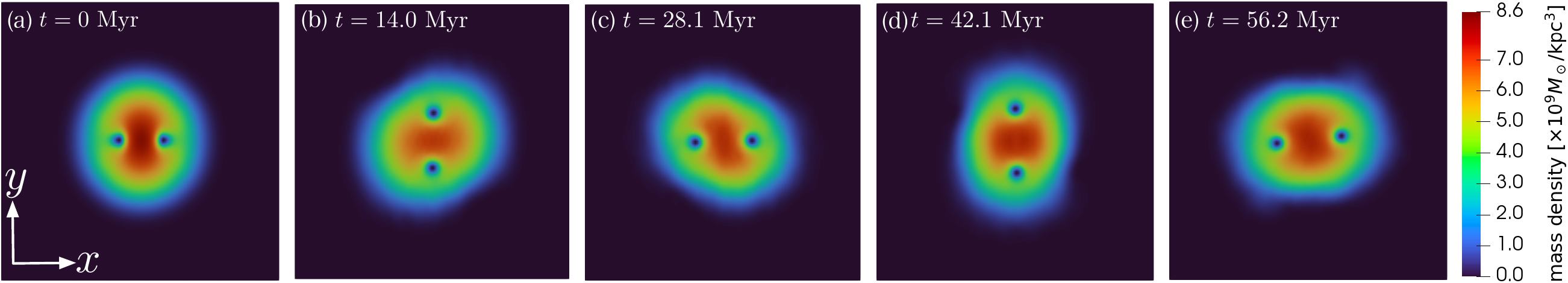}     
\caption{
The mass density profiles in the plane $z=0$ when $r_o\approx2.0$ kpc.
(a) $0~\mathrm{Myr}$, (b) $14.0~\mathrm{Myr}$, (c) $28.1~\mathrm{Myr}$, (d) $42.1~\mathrm{Myr}$, and (e) $56.2~\mathrm{Myr}$.
The box size is approximately $25.8$ kpc.
}
\label{figure1}
\end{figure*}
\subsection{NON-CIRCULAR ORBITAL ROTATION OF TWO PARALLEL QUANTIZED VORTICES}
Even in the self-gravitating BEC, two parallel quantized vortices co-rotate around the center of the BEC similarly to those in the conventional BEC as shown in Fig.\ref{figure1} and a corresponding video in the Supplemental Material
\footnote{
See Supplemental Material for the time evolution of the mass density profile in the plane $z=0$ when $r_o\approx2.0$ kpc.
}.
It shows the mass density profiles in the plane $z=0$ at various time when $r_o\approx2.0$ kpc.
Two dark-colored circles show the quantized vortices.
Initially positioned on the $x$-axis as depicted in Fig. \ref{figure1}(a), they rotate at a constant angular velocity, completing a full revolution around $56.2$ Myr, as illustrated in Fig. \ref{figure1}(e).
Note that Fig. \ref{figure1} also shows the deformation of the BEC, which is analyzed later.

\begin{figure}
\centering
\includegraphics[width=8.5cm]{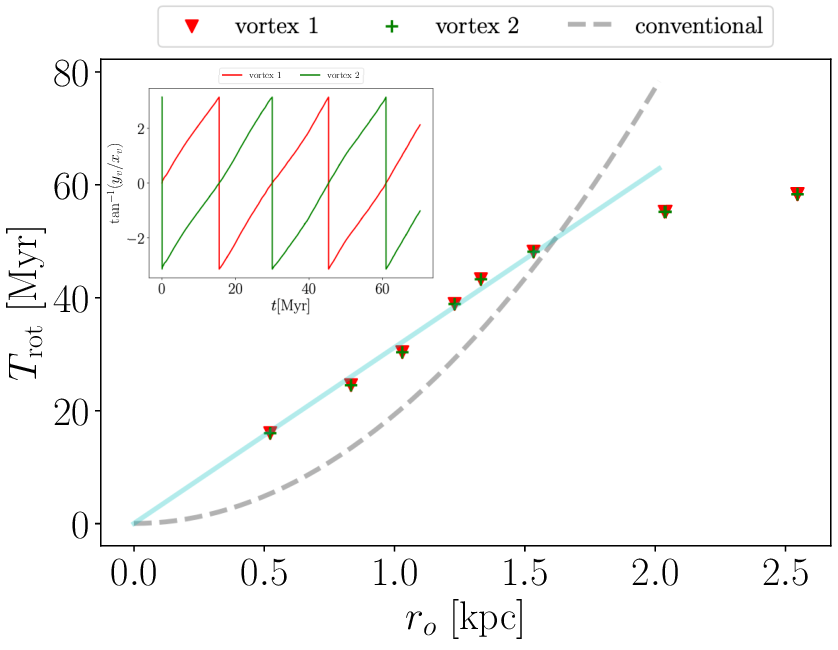}     
\caption{
The relationship between the rotational period $T_\mathrm{rot}$ and $r_o$, compared to that of a conventional BEC given by Eq. (\ref{period_uniform}).
The inset is the time evolution of the rotational angle of two quantized vortices in the plane $z=0$ when $r_o\approx1.0$ kpc.
}
\label{figure2}
\end{figure}
The rotation of two quantized vortices at a constant angular velocity is seen in the inset of Fig. \ref{figure2}.
It is the time evolution of the rotational angle of two quantized vortices $\tan^{-1}(y_{v,1}/x_{v,1})$ and $\tan^{-1}(y_{v,2}/x_{v,2})$ when $r_o\approx1.0$ kpc.
These rotational angles linearly increase with the same tendency, ensuring both vortices rotate at the same constant angular velocity.
Even though $r_o$ changes, the quantized vortices exhibit such a linear increase of the rotational angle.

Although the particular angular velocity or period characterizes the orbital rotation of two quantized vortices in the self-gravitating BEC similar to that in the usual BEC, its $r_o$-dependence is different as shown in Fig. \ref{figure2}. 
In this figure, both red triangles and green crosses are not put on a gray dashed line, which shows the rotational period $T_\mathrm{rot}$ does not quadratically increase with $r_o$.
Particularly, in $r_o\lesssim1.5$ kpc they are put on a cyan straight line, and namely $T_\mathrm{rot}\propto r_o$.
This differs from the rotational period of two vortices in a conventional BEC, namely $T_\mathrm{rot}=(8\pi^2/\kappa)r_o^2$ of Eq. (\ref{period_uniform}) which is derived in Appendix A, shown by the black dashed line.
The extension of the period suggests the presence of an additional drag force acting on the vortices in the self-gravitating BEC.
We infer that this drag force originates from gravitationally long-range interactions, analogous to the acceleration of the rotation of quantized vortices in a dipolar BEC \cite{zhao2021, zhao2022}.
On the other hand, when $r_o\gtrsim2.0$ kpc, the rotational period does not increase significantly even if $r_o$ increases.
We predict that it is caused by the finite size effect analogous to two vortices in a uniform BEC in a cylindrical container as shown in Appendix A.

\begin{figure}
\centering
\includegraphics[width=8cm]{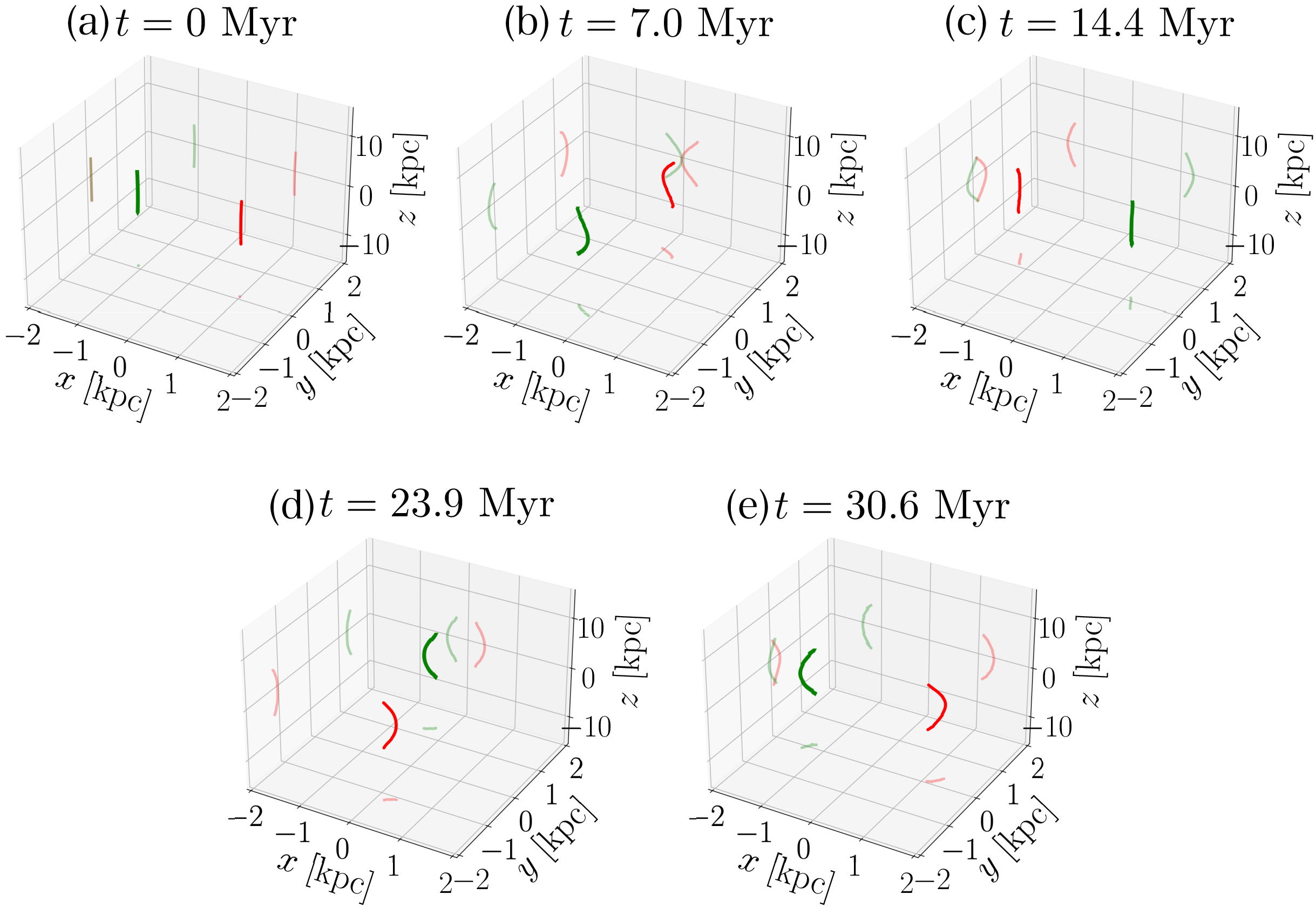}     
\caption{
The vortex cores in the three-dimensional system at (a) $0$ Myr, (b) $7.0$ Myr, (c) $14.4$ Myr, (d) $23.9$ Myr, and (e) $30.6$ Myr with their projections in $xy$, $yz$, and $zx$ planes when $r_o\approx1.0$ kpc.
The red and green lines show the vortex cores.
}
\label{figure3}
\end{figure}
In the self-gravitating BEC, two quantized vortices display the bending of the vortex cores with corotating.
Figure \ref{figure3} and a corresponding video in the Supplemental Material \footnote{
See Supplemental Material for the time evolution of the vortex cores in the three-dimensional system when $r_o\approx1.0$ kpc.
} show the time evolution of two vortex cores when $r_o\approx1.0$ kpc in the three-dimensional system.
Two vortex cores initially show straight lines along the $z$-axis in Fig. \ref{figure3} (a).
As the cores start to rotate, they undergo deformation, altering their bending direction as shown in Figs. \ref{figure3} (b)-(e).
\begin{figure}
\centering
\includegraphics[width=8cm]{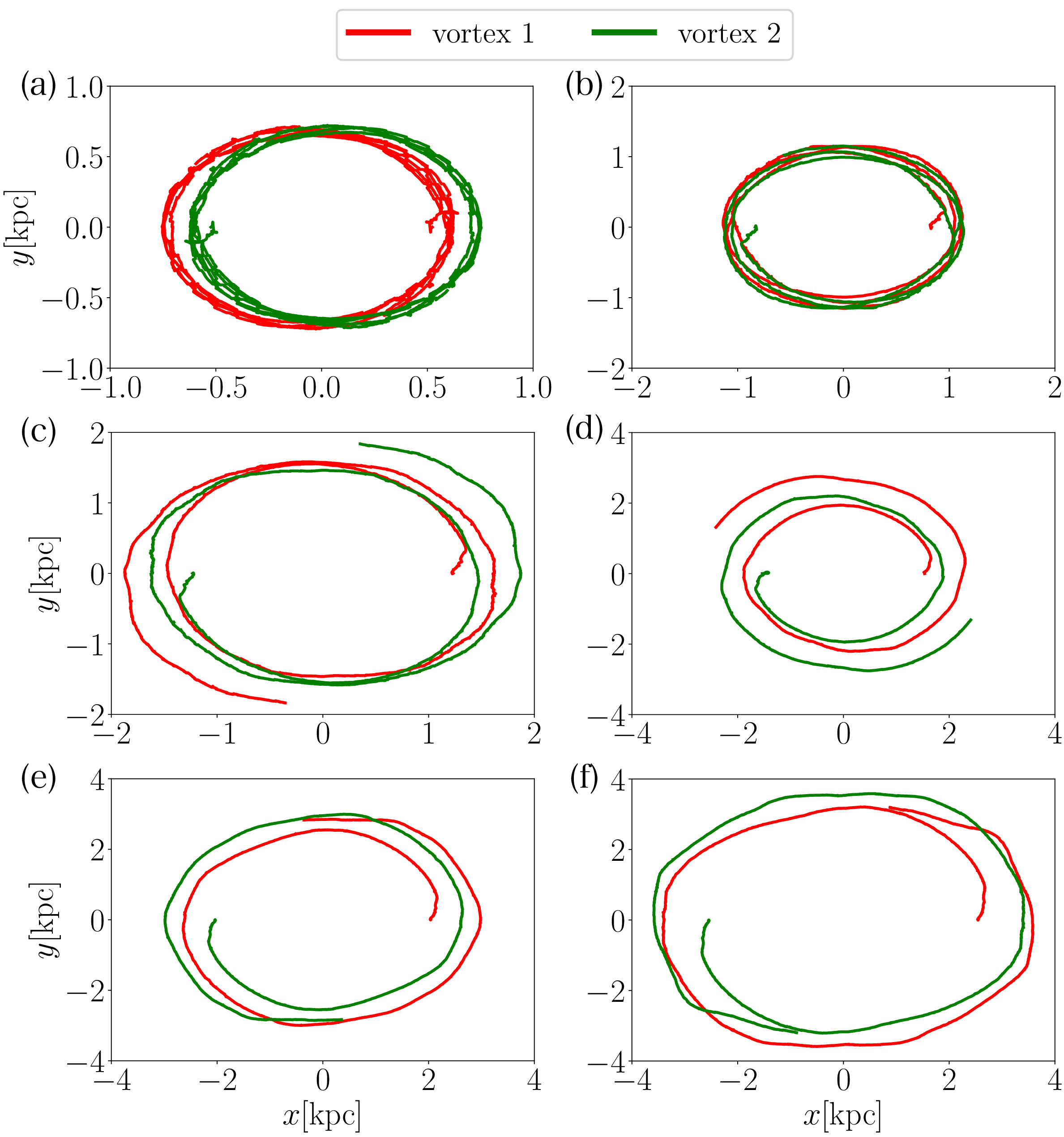}     
\caption{
The trajectories of $(x_{v,1},y_{v,1})$ (red line) and $(x_{v,2},y_{v,2})$ (green line) shown by the orbital rotation of the quantized vortices.
(a) $r_o\approx0.5$ kpc, (b) $r_o\approx0.8$ kpc, (c) $r_o\approx1.2$ kpc, (d) $r_o\approx1.5$ kpc, (e) $r_o\approx20$ kpc, and (f) $r_o\approx2.5$ kpc.
}
\label{figure4}
\end{figure}
Then, the vortex cores trace a non-circular trajectory as shown in Figs. \ref{figure4} (a)-(f).
It shows the time evolution of the vortex positions in the $z=0$ planes $(x_{v,1},y_{v,1})$ and $(x_{v,2},y_{v,2})$ as changing $r_o$.
The behavior can be divided into two types whether $r_o\lesssim1.0$ kpc in Figs. \ref{figure4} (a) and (b) or $r_o\gtrsim1.2$ kpc in Figs. \ref{figure4} (c)-(f).
In the former cases, the vortex cores follow elliptical orbits, whereas in the latter, they spiral outward with fluctuations.

First, let us examine the elliptical orbits in detail when $r_o$ is smaller than $1.0$ kpc.
We focus on two important terms; one is whether two orbits overlap or not, and the other is whether each orbit is closed or not.
In Fig. \ref{figure4} (a) corresponding to the case of $r_o\approx0.5$ kpc, the elliptical trajectories just shift from each other.
Conversely, as shown in Fig. \ref{figure4} (b), for other cases such as $r_o\gtrsim0.8$ kpc, the two ellipses overlap.
Additionally, as $r_o$ increases, the rotational orbits deviate from a perfectly elliptical shape, and as a result, the orbits are not strictly closed when $r_o\gtrsim0.8$ kpc.
Such non-closed elliptical trajectories are also observed in a BEC with dipole-dipole interactions \cite{zhao2021}.
Hence, long-range interactions in a BEC may distort the trajectory of the corotating vortices from a closed circular form.

\begin{figure}
\centering
\includegraphics[width=8cm]{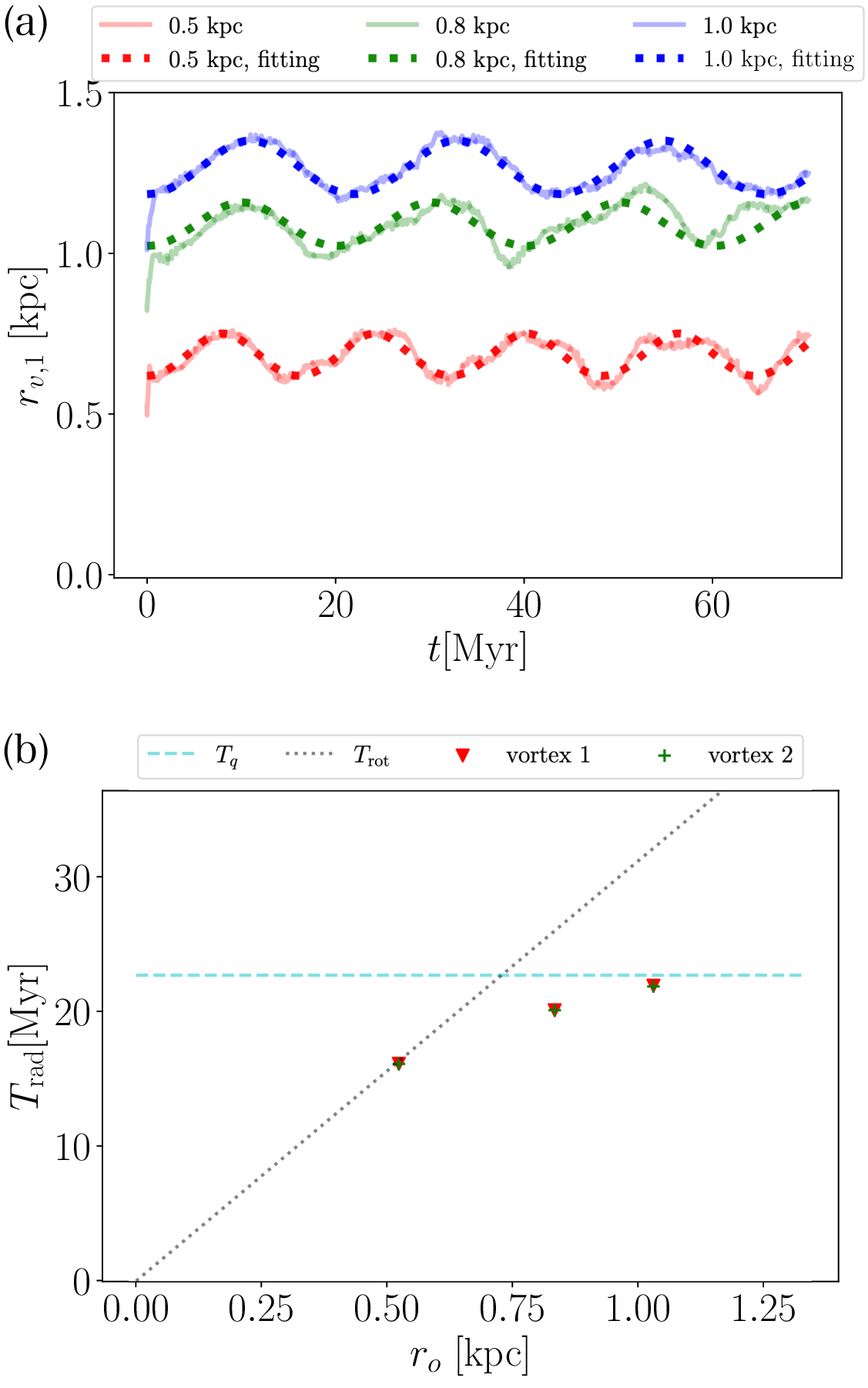}     
\caption{
(a) The time evolution of the distance of vortex 1 from the center of the BEC $r_{v,1}$ and the fitting results using the polar form of the ellipse equation.
(b) The relationship between the initial distance of the vortex from the center of the BEC $r_{o}$ and the period of the oscillation estimated by fitting $T_\mathrm{rad}$, and the comparison to the rotational period of the vortex $T_\mathrm{rot}$ and that of the quadrupole mode without the quantized vortex $T_\mathrm{q}$.
}
\label{figure5}
\end{figure}
Figure \ref{figure5} presents the quantitative analysis of the distance of each vortex from the center of the BEC $r_{v,l}=\sqrt{x_{v,l}^2+y_{v,l}^2}$, where $l$ is $1$ or $2$, revealing the characteristics of elliptical orbits.
As depicted in Fig. \ref{figure5} (a), the quantized vortex exhibits radial oscillations.
By fitting the data with the polar form of the ellipse equation, expressed as
\be
r_{v,l}
=
\f{A(1-\epsilon^2)}{1-(-1)^l\epsilon\cos\left(\f{2\pi}{T_\mathrm{rad}}t\right)},
\label{ellipse_equation}
\ee
all results align with our numerical findings.
Here, $A$ represents the semi-major axis, $\epsilon$ denotes the eccentricity, and $T_\mathrm{rad}$ is the period of the radial oscillation.
This confirms quantitatively that the rotational orbits exhibit an elliptical form due to the radial oscillation of the quantized vortex.
Additionally, the overlap of two ellipses and the non-closed trajectory can be explained by the eccentricity and the oscillatory period as follows.
The eccentricity reaches a maximum of $\epsilon\approx0.09$ when $r_o\approx0.5$ kpc and decreases to $\epsilon\approx0.06$ when $r_o\gtrsim0.8$ kpc, leading to the overlap of the orbits of co-rotating two vortices.
According to Fig. \ref{figure5} (b), which illustrates the $r_o$-denpendence of $T_\mathrm{rad}$, it is comparable to the rotational period $T_\mathrm{rot}$ when $r_o\approx0.5$ kpc, but increases with $r_o$, approaching a particular value.
This increase in $T_\mathrm{rad}$ relative to $T_\mathrm{rot}$ as $r_o$ causes the rotational trajectory to become non-closed, due to phase discrepancies between the orbital rotation and the radial oscillation. 
Note that, as demonstrated later, the value toward which $T_\mathrm{rad}$ converges with increasing $r_o$ is estimated to correspond to the period $T_q$ of the quadrupole mode in the self-gravitating BEC without vortices.

Next, we conduct a detailed analysis of the spiraling behavior of two quantized vortices when $r_o\gtrsim1.2$ kpc.
Figure \ref{figure4} (c) demonstrates that the rotational trajectories of the two quantized vortices exhibit a time-dependent transition from elliptical to spiral when $r_o\approx1.2$ kpc; $r_{v,1}$ and $r_{v,2}$ initially oscillate until a certain point in time, after which they begin to increase while still oscillating.
Such a time-dependent switching is absent when $r_o\approx1.5$ kpc, as seen in Fig. \ref{figure4} (d), where the two quantized vortices simply spiral outside while fluctuating.
More intriguingly, for the cases where $r_o\approx2.0$ kpc in Fig. \ref{figure4} (e) and $2.5$ kpc in Fig. \ref{figure4} (f), two quantized vortices, after spiraling out sufficiently, then begin to spiral in.

\begin{figure}
\centering
\includegraphics[width=8cm]{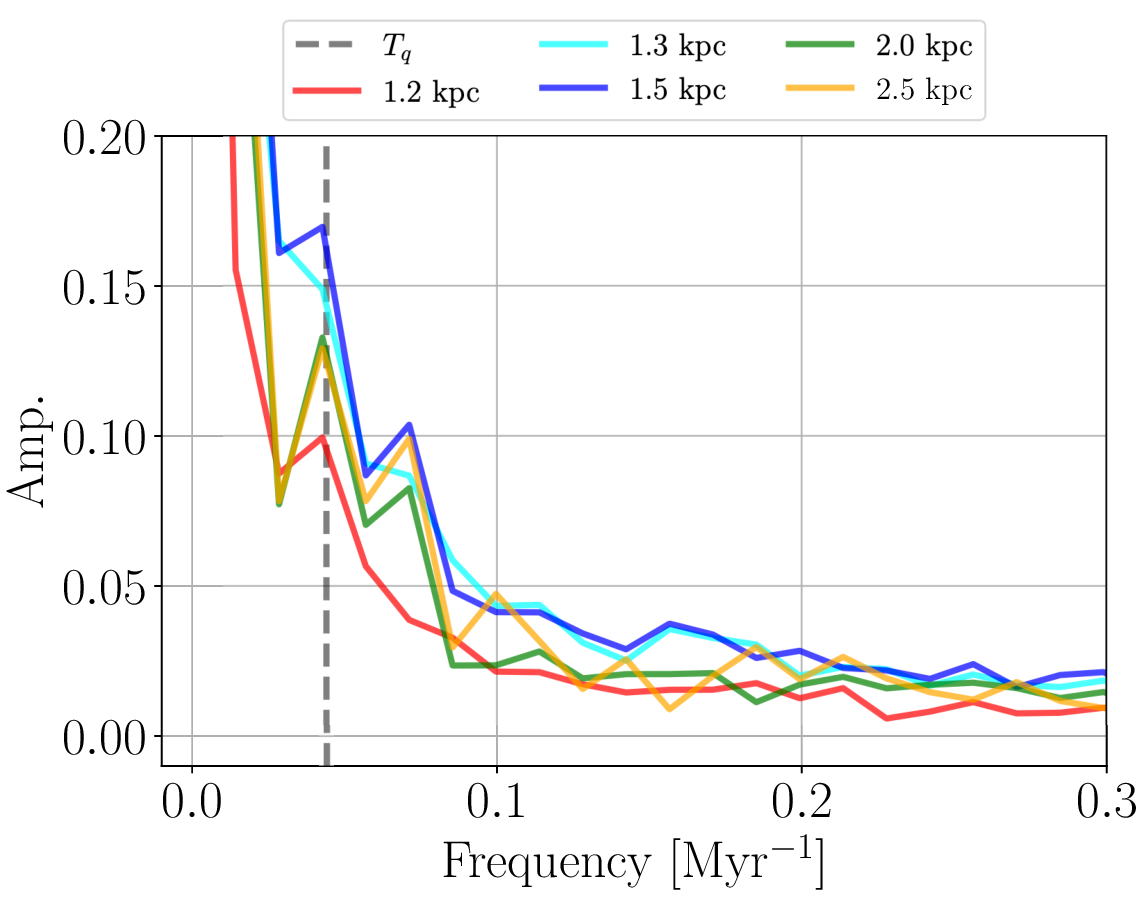}     
\caption{
The applications of the vortex-point model considering the two same-charged quantized vortices in a circular container to our numerical results.
(a) The time evolution of a function $Y[r_{v,l}]$ which is explicitly written by Eq. (\ref{function_y}).
The solid lines show the numerical results obtained by substituting $r_{v,l}$ into Eq. (\ref{function_y}), and the dotted lines demonstrate the linear fitting with those to estimate the mean value of $\alpha$ in Eq. (\ref{vp_result}).
(b) The parameter $\alpha$ which shows the effective dissipation of the quantized vortex vs the initial distance of the quantized vortex from the center of the BEC, as indicated by black circles.
The cyan line shows the coherence length $\xi$ and the magenta line indicates twice that.
}
\label{figure7}
\end{figure}
These spiral trajectories or increases of $r_{v,l}$ during rotation suggest the presence of dissipations acting on the quantized vortices, despite the fact that we consider the non-dissipative GPP model (see Eqs. (\ref{gp}) or (\ref{dimensionless_gp})).
We employ the vortex-point model with an effective dissipation analogous to mutual friction as a simplified approach to estimate the magnitude of these dissipative effects.
For simplicity, we consider that two quantized vortices, with identical charges, are symmetrically positioned within a circular container of radius $R$.
In this situation, the time evolution of the distance of each quantized vortex from the center of the BEC satisfies the following relation:
\be
\alpha t=Y[r_{v,l}]
\label{vp_result}
\ee
as derived in Appendix A.
Here, the parameter $\alpha$ quantitatively expresses the effective dissipation, and the function $Y[r_{v,l}]$ is defined as
\be
\begin{split}
\f{\kappa}{2\pi}Y[r_{v,l}]
\equiv
& \f{4\sqrt{3}}{9}R^2
\tan^{-1}\left(\f{\sqrt{3}r_{v,l}^2}{R^2}\right) \\
&-
\f{4\sqrt{3}}{9}R^2
\tan^{-1}\left(\f{\sqrt{3}r_{o}^2}{R^2}\right) \\
&-
\f{r_{v,l}^2-r_o^2}{3}.
\end{split}
\label{function_y}
\ee
Substituting the numerical results of $r_{v,l}$ into Eq. (\ref{function_y}), the time evolution of $Y[r_{v,l}]$ can be obtained, as exhibited by solid curves in Fig. \ref{figure7} (a).
This figure allows us to categorize these solid curves into two groups; $Y[r_{v,l}]$ hardly increases when $r_o$ is small while it increases linearly, and dissipation works when $r_o$ is large.
The mean increasing rate of $Y[r_{v,l}]$ corresponds to the magnitude of effective dissipation $\alpha$, as indicated in Eq (\ref{vp_result}).
By estimating the values of $\alpha$ through linear fitting as shown by the dotted lines in Fig. \ref{figure7} (a), we observe the $r_o$-dependence in Fig. \ref{figure7} (b).
The rapid increase between $\xi$ and $2\xi$ indicates that the critical distance at which the rotational trajectories transit from elliptical to spiral is comparable to the vortex core size.
Therefore, we infer that inter-vortex interactions are closely related to the emergence of effective dissipation.
It is noteworthy that effective dissipation also depends on time.
For instance, when $r_o\approx2.0$ kpc and $2.5$ kpc, $Y[r_{v,l}]$ saturates around $t=40$ Myr, and continues to fluctuate thereafter.
Moreover, when $r_o\approx1.2$ kpc and $1.3$ kpc, the time-dependent transition of the trajectories from ellipses to spirals likely alters the magnitude of effective dissipation.
However, in this study, we do not explore the time dependence of the effective dissipation in further detail.

\begin{figure}
\centering
\includegraphics[width=8cm]{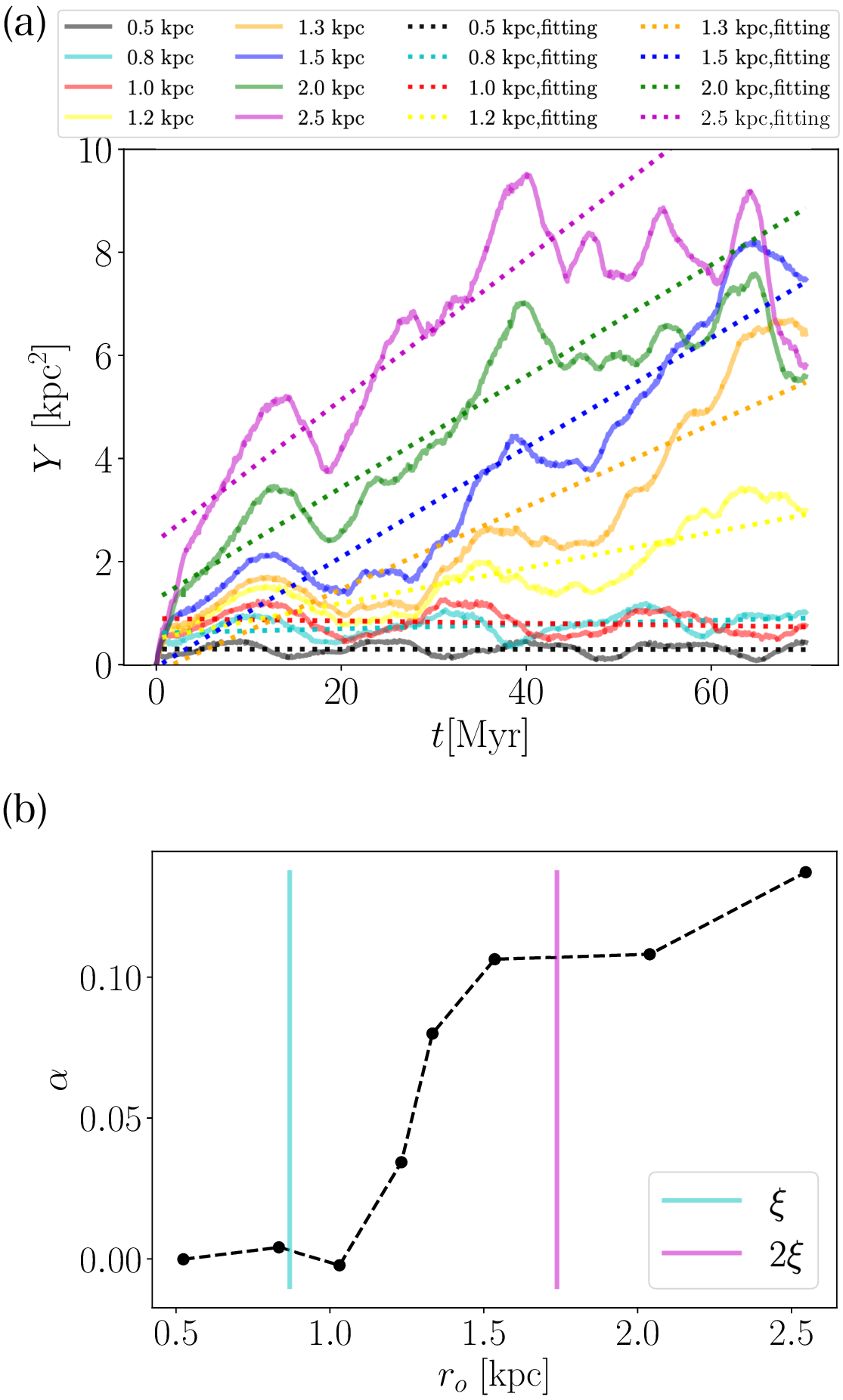}     
\caption{
The spectrum of $r_{v,1}$ using the FFT analyses when $r_o\gtrsim1.2$ kpc, namely the trajectory draws spiral.
The solid lines respectively show the corresponding numerical results while the black dashed line is the frequency of the quadrupole mode in a self-gravitating BEC with no quantized vortex, represented as Eq. (\ref{Tq}).
}
\label{figure6}
\end{figure}
While the spiral behavior of the two quantized vortices varies with $r_o$, as shown in Figs. \ref{figure4} (c)-(f), FFT analysis of the time evolution of $r_{v,l}$ reveals that their radial fluctuations share a common frequency.
Figure \ref{figure6} presents the spectra of $r_{v,1}$ when $r_o\gtrsim1.2$ kpc, obtained from the FFT analysis.
These spectra consistently exhibit a peak corresponding to the quadrupole mode in a self-gravitating BEC without vortices, whose period is given by
\be
T_\mathrm{q}
=
2\pi
\sqrt{
\f{5\pi(\pi^2-6)}{3}
\sqrt{\f{\hbar^{6}a^{3}}{G^{5}m^{9}}}
}
\f{1}{\sqrt{M}}
\label{Tq}
\ee 
assuming the total mass $M$ is sufficiently large \cite{giovanazzi2001, asakawa2024}.
Hence, $r_{v,l}$ oscillates with a frequency comparable to that of the quadrupole mode even though $r_o$ just changes in the situations where the quantized vortices spiral outward.
We can conclude that, as $r_o$ increases, the period of the radial oscillation $T_\mathrm{rad}$ becomes smaller than the rotational one $T_\mathrm{rot}$ and approaches $T_\mathrm{q}$.

\begin{figure*}
\centering
\includegraphics[width=18cm]{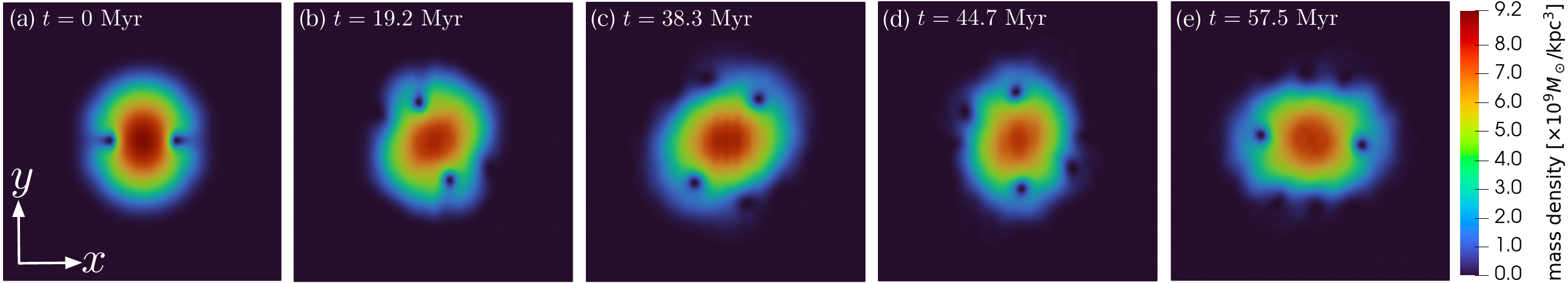}     
\caption{
The mass density profiles in the plane $z=0$ when $r_o\approx3.0$ kpc.
(a) $0~\mathrm{Myr}$, (b) $19.2~\mathrm{Myr}$, (c) $38.3~\mathrm{Myr}$, (d) $44.7~\mathrm{Myr}$, and (e) $57.5~\mathrm{Myr}$.
The box size is approximately $25.8$ kpc.
}
\label{figure8}
\end{figure*}
In all the figures presented above, the numerical results for cases where $r_o\approx3.0$ kpc and $4.0$ kpc are not contained.
This is due to the unexpected emergence of the ghost vortices near the edge of the BEC, which hinders the accurate determination of the coordinates of the two quantized vortices.
Figures \ref{figure8} (a)-(e) and a corresponding video in the Supplemental Material \footnote{
See Supplemental Material for the time evolution of the mass density profile in the plane $z=0$ when $r_o\approx3.0$ kpc.
} depict the time evolution of the mass density profiles, sliced at the $z=0$ plane, where $r_o\approx3.0$ kpc.
In the early stage demonstrated in Fig. \ref{figure8} (b), surface fluctuations induced by the rotation of the quantized vortices cause partial concavities at the edge of the BEC, compared to the initial profile in Fig. \ref{figure8} (a).
These concavities evolve into two ghost vortices in Fig. \ref{figure8} (c), and their number gradually increases to 4-6, as illustrated in Figs. \ref{figure8} (d) and (e). 
Furthermore, according to Fig. \ref{figure8} (e), two quantized vortices complete approximately one rotation at $t\approx58$ Myr when $r_o\approx3.0$ kpc.
This rotational period is shorter than that for $r_o\approx2.5$ kpc, as shown in Fig. \ref{figure2}, supporting the probability that finite-size effects accelerate the rotational speed of the quantized vortices, even in a self-gravitating BEC.

\subsection{DEFORMATION, ROTATION, AND OSCILLATION OF THE BEC}
Figure \ref{figure1} and Figure \ref{figure8} also demonstrate that as the two quantized vortices corotate, the BEC deforms into an ellipsoidal shape and rotates around the $z$-axis.
This ellipsoidal deformation and rotation are quantitatively analyzed by tracking the effective sizes in each direction $X(t)$, $Y(t)$, and $Z(t)$ defined as
\be
X(t)
=
\sqrt{
\f{m}{M}
\int d\mathbf{r}
x^2
|\psi(\bm{r},t)|^2
},
\label{effective_semiaxis_x}
\ee
\be
Y(t)
=
\sqrt{
\f{m}{M}
\int d\mathbf{r}
y^2
|\psi(\bm{r},t)|^2
},
\label{effective_semiaxis_y}
\ee
and
\be
Z(t)
=
\sqrt{
\f{m}{M}
\int d\mathbf{r}
z^2
|\psi(\bm{r},t)|^2
}.
\label{effective_semiaxis_z}
\ee
In our numerical results, the time evolutions of these semi-axes exhibit complex oscillations, indicating that the superposition of collective modes drives the deformation and rotation within the BEC.

\begin{figure}
\centering
\includegraphics[width=8.5cm]{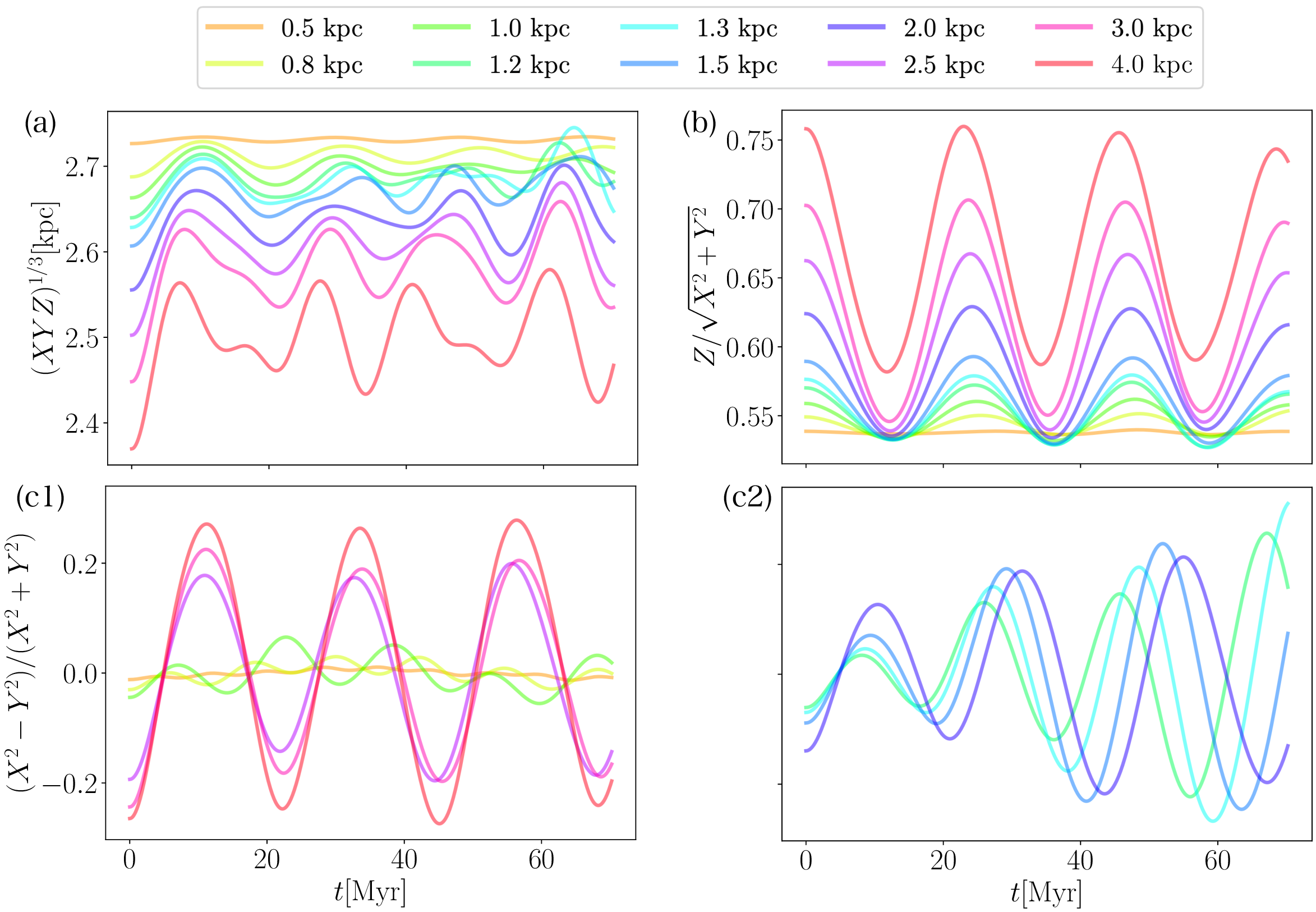}     
\caption{
Time evolutions of quantities that characterize an ellipsoidal shape of the BEC when $r_o$ changes.
(a) the average radius $(XYZ)^{1/3}$, (b) the aspect of the BEC along the $z$-axis $Z/\sqrt{X^2+Y^2}$, (c1) and (c2) the aspect of the BEC in the $xy$ planes $(X^2-Y^2)/(X^2+Y^2)$.
}
\label{figure9}
\end{figure}
We focus on three collective modes: the spherical breathing (monopole) mode, the quadrupole mode along the $z$-axis, and the quadrupole mode in the $xy$ plane.
These collective modes are characterized by specific quantities of the ellipsoidal shape: the mean radius $(XYZ)^{1/3}$ and two aspect ratios of the BEC $Z/\sqrt{X^2+Y^2}$ and $(X^2-Y^2)/(X^2+Y^2)$.
As shown in Fig. \ref{figure9}, the time evolution of these quantities exhibits different oscillations.
These figures reveal that each oscillation has a different dependence on $r_o$.
As $r_o$ increases, the oscillation of the mean radius results from the superposition of several modes, the oscillation of $Z/\sqrt{X^2+Y^2}$ remains monotonic and independent of $r_o$, and the oscillation of $(X^2-Y^2)/(X^2+Y^2)$ converges to a specific mode.
According to Fig. \ref{figure9} (a), the size of the BEC decreases with increasing $r_o$ 
\footnote{
The reduction in the size of the BEC is attributed to stronger gravitational binding resulting from the increased central mass density. 
When two quantized vortices are positioned far from the center of the BEC, they do not interfere with the accumulation of particles in the central region. 
Consequently, the central density rises with increasing $r_o$, thereby enhancing the gravitational potential.
}.
On the other hand, Figs. \ref{figure9} (b), (c1), and (c2) demonstrate that as $r_o$ increases, the initial configuration becomes distorted from the spherical shape, leading to an amplification of the oscillations in both aspect ratios.
The most intriguing aspect is that, as shown in Fig. \ref{figure9} (c2), is the gradual increase in the amplitude of $1.2~\textrm{kpc}\lesssim r_o\lesssim 2.0~\textrm{kpc}$, despite the system being non-dissipative.
This suggests that the quadrupole mode in the $xy$ plane contributes to the effective dissipation of the corotating quantized vortex.

\begin{figure}
\centering
\includegraphics[width=8.5cm]{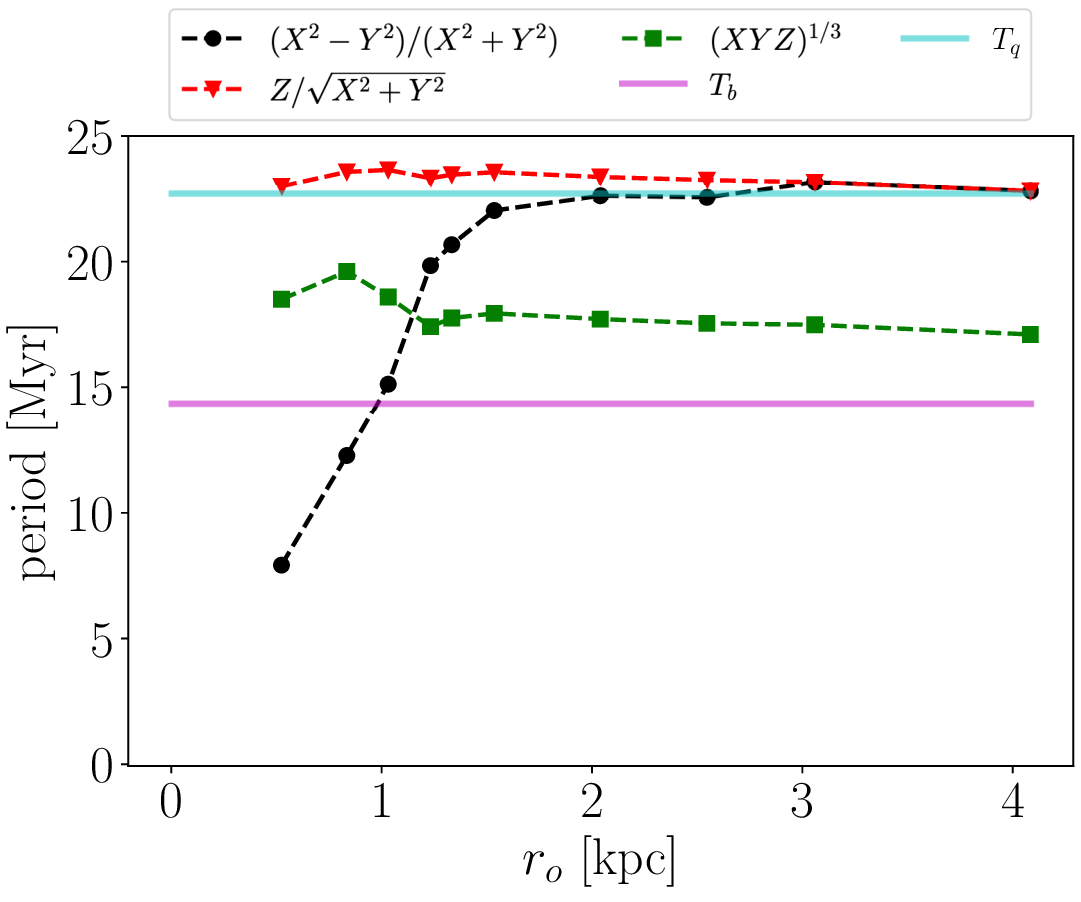}     
\caption{
$r_o$-dependence of the periods corresponding to a maximum peak of spectra of each oscillation (black circles: the mean radius, red triangles: the aspect ratio of the BEC along the $z$-axis, green squares: the aspect ratio of the BEC in the $xy$ plane).
The solid lines show the period of collective modes in a self-gravitating BEC without quantized vortices: magenta represents the breathing mode and cyan represents the quadrupole mode.
}
\label{figure10}
\end{figure}
As observed in Fig. \ref{figure9}, the presence of quantized vortices also affects the frequency of these collective modes, particularly the quadrupole mode in the $xy$ plane. 
Figure \ref{figure10} shows the maximum spectra peaks obtained by the FFT analysis of the oscillations in Fig. \ref{figure9}, illustrating the relationship between $r_o$ and the periods of each collective mode.
The breathing mode shows that, while its period remains largely unaffected by $r_o$, the presence of quantized vortices causes a shift in its frequency from that of the no-vortex case, which is given by
\be
T_\mathrm{b}
=
2\pi
\sqrt{
\f{2\pi(\pi^2-6)}{3}
\sqrt{\f{\hbar^{6}a^{3}}{G^{5}m^{9}}}
}
\f{1}{\sqrt{M}}
\label{Tb}
\ee
\cite{giovanazzi2001, asakawa2024}.
This increase in the period due to the presence of quantized vortices qualitatively aligns with the analytical results concerning a quantized vortex at the center of a self-gravitating BEC \cite{ghosh2002}.
While the quadrupole mode along the $z$-axis remains unaffected by the quantized vortices, the quadrupole mode in the $xy$-plane exhibits an increase in its period with $r_o$, approaching $T_q$ expressed in Eq. (\ref{Tq}).
We infer that the $r_o$-dependence of the quadrupole mode in the $xy$ plane is linked to the rotation of the BEC, caused by the splitting of two degenerate frequencies due to the presence of the quantized vortices.
It is well-known that such rotation occurs when a quantized vortex is situated at the center of a trapped BEC, and its angular velocity is determined by the difference between the two shifted frequencies \cite{stringari}.

As mentioned above, the oscillations of the mean radius and the aspect ratio in the $xy$ plane are superimposed across several modes depending on $r_o$. 
In particular, the latter is likely attributable to the lifting of the degeneracy in the quadrupole mode. 
However, we were unable to detect these frequencies in the present study, probably because the lower split frequency does not lie within the time window covered by our numerical calculations. 
Further investigations will be pursued as part of our future work.

\subsection{ENERGY TRANSFER TRIGGERED BY RESONANCE BETWEEN QUANTIZED VORTICES AND QUADRUPOLE MODE}
\begin{figure}
\centering
\includegraphics[width=8cm]{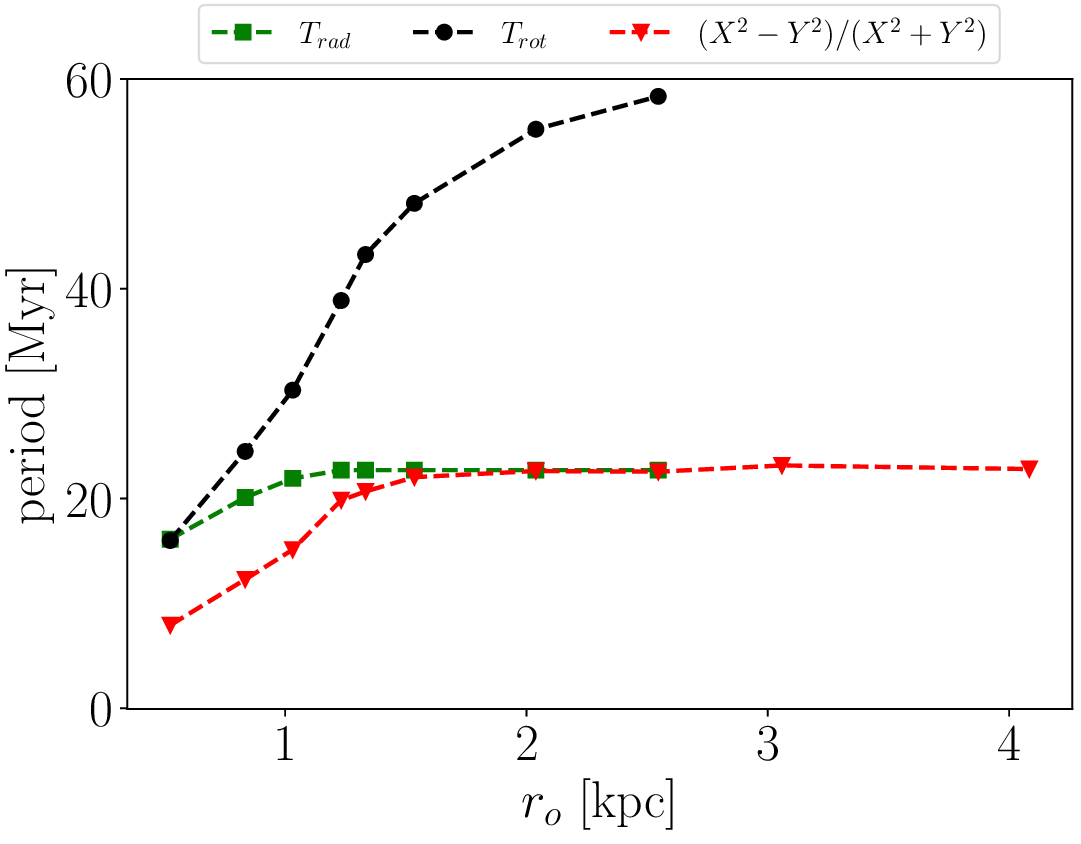}     
\caption{
The relationship between $r_o$ and three periods.
Black circles show the period rotating two quantized vortices $T_\mathrm{rot}$, green squares show that oscillating in the radial direction $T_\mathrm{rad}$, and red triangles show that of $(X^2-Y^2)/(X^2+Y^2)$, namely the period of the quadrupole mode in the $xy$ plane.
}
\label{figure11}
\end{figure}
When the two corotating quantized vortices trace spiral trajectories, their radial oscillations resonate with the quadrupole mode of the BEC.
As shown in Fig. \ref{figure11}, $T_\mathrm{rad}$ increases with $r_o$ and gradually approaches the period of $(X^2-Y^2)/(X^2+Y^2)$.
As a result, when $r_o\gtrsim1.2$ kpc, or two quantized vortices spiral outward, the radial oscillation of the quantized vortices has a similar period to the oscillation of the aspect ratio $(X^2-Y^2)/(X^2+Y^2)$, corresponding to the quadrupole mode in the $xy$ plane.
Thus, when two quantized vortices rotate in this system, their radial fluctuations couple with the quadrupole mode.

\begin{figure}
\centering
\includegraphics[width=8.5cm]{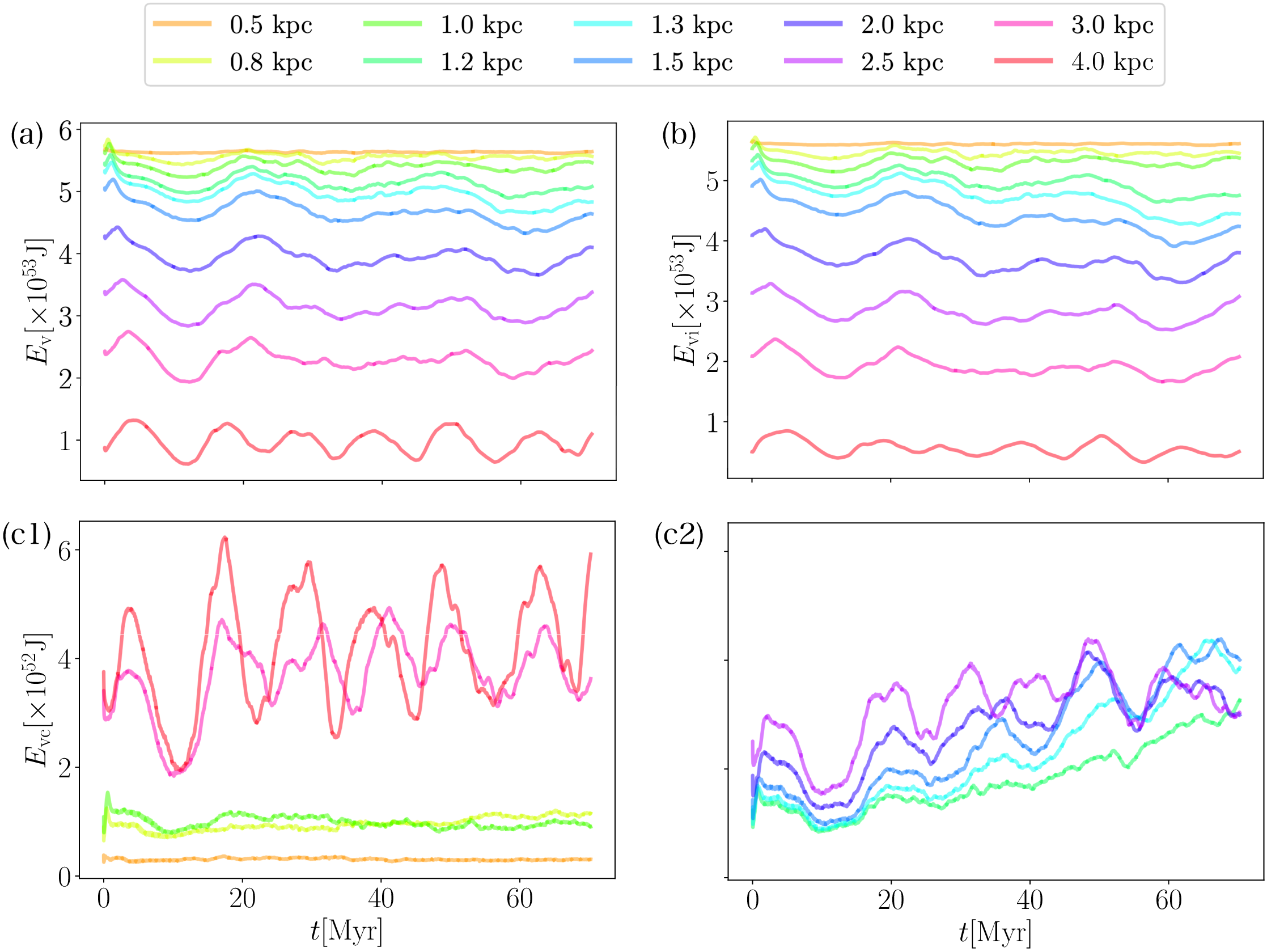} 
\caption{
Time evolutions of kinetic energy components when $r_o$ changes.
(a) the velocity-field component $E_\mathrm{v}(t)$, (b) incompressible part $E_\mathrm{vi}(t)$, (c1) and (c2) compressible part $E_\mathrm{vc}(t)$.
}
\label{figure12}
\end{figure}
Figure \ref{figure12} suggests that the effective dissipation of the quantized vortices is associated with its resonance with the quadrupole mode.
As shown in Figs. \ref{figure12} (a) and (b), the incompressible part $E_\mathrm{vi}(t)$ defined as Eq. (\ref{incompressible_energy}) predominantly governs the velocity-field contribution of the kinetic energy $E_\mathrm{v}(t)$.
However, in Figs. \ref{figure12} (b), $E_\mathrm{vi}(t)$ exhibits a slight decrease when $1.2~\textrm{kpc}\lesssim r_o \lesssim 2.5~\textrm{kpc}$ where resonance between the quantized vortices and the quadrupole mode in the $xy$ plane occurs.
Since the system conserves the total energy, this reduction suggests that the energy of the quantized vortices is gradually redistributed into other energy components during their corotation. 
In contrast, the time evolution of the compressible part $E_\mathrm{vc}(t)$ in Figs. \ref{figure12} (c1) and (c2) shows distinct qualitative changes depending on whether $1.2~\textrm{kpc}\lesssim r_o \lesssim 2.5~\textrm{kpc}$ or not.
The former case in Fig. \ref{figure12} (c2) shows $E_\mathrm{vc}(t)$ increases with fluctuating and saturates in late time, which implies that phonons are excited in the BEC until saturation.
On the other hand, the latter one in Fig. \ref{figure12} (c1) does not cause the rise of $E_\mathrm{vc}(t)$.
Hence, oscillations or deformations of the BEC are time-dependently amplified when the effective dissipation works in the quantized vortices, likely caused by the energy transfer by phonons from the incompressible part.
Such time-dependent amplification of oscillation is also shown by $(X^2-Y^2)/(X^2+Y^2)$, and therefore we conclude that the quadrupole mode in the $xy$ plane receives the energy from vortices through emitted phonons.
Notably, the time evolution of the gravitational component of Eq. (\ref{gravitational_energy}) and the contact-interaction component of Eq. (\ref{contact_energy}) exhibit the complicated oscillation similar to the mean radius of the BEC as seen in Fig. \ref{figure9} (a), which reflects that the configuration is composed of the competition between two self-interactions.

Figure \ref{figure12} (a) is likely to indicate that the corotation of the two quantized vortices becomes coupled with the quadrupole mode in the $xy$ plane when $r_o$ becomes sufficiently large.
For cases where $r_o\gtrsim3.0~\textrm{kpc}$, the time evolution of $E_\mathrm{vc}(t)$ exhibits more pronounced oscillations compared to when $r_o\lesssim2.5~\textrm{kpc}$, yet the increase observed in Fig. \ref{figure12} (b) does not manifest. 
We infer that the rotating quantized vortices near the BEC surface interfere with the quadrupole mode in the $xy$ plane, making it challenging to discern whether the deformation and rotation are primarily driven by the quadrupole mode or the quantized vortices.

\section{CONCLUSIONS AND DISCUSSIONS}
We numerically investigate the corotation of two parallel quantized vortices in a self-gravitating BEC using the three-dimensional GPP equation. 
The gravitational potential that confines the BEC is influenced by the motion of the BEC and the presence of quantized vortices through its density dependence. 
As a result, the dynamics of quantized vortices become complex and highly intriguing. 
This work seeks to elucidate the effects of gravitational interactions on the motion of quantized vortices within the BEC and to uncover non-trivial quantum hydrodynamical phenomena driven by self-gravity.

The gravitational interaction exerts a drag force on the rotating quantized vortices, thereby prolonging their rotational period. 
When the two quantized vortices are initially positioned near the center of the BEC, their rotational period increases linearly with their initial distance from the center, differing from the quadratic increase observed in a uniform BEC. 
This difference suggests that the gravitational interaction acts as a drag force on the co-rotating vortices and reduces their rotational velocity. 
Conversely, when the vortices are initially positioned near the surface of the BEC, the rotational period does not increase significantly due to finite-size effects.

The corotating quantized vortices in the self-gravitating BEC exhibit radial oscillations. 
As the initial distance of the vortices from the center increases, this oscillation period extends and converges towards that of the quadrupole mode in a self-gravitating BEC without quantized vortices. 
Compared to the rotational period, the period of radial oscillations aligns when the two vortices are sufficiently close but becomes relatively shorter as they are further separated. 
The resulting discrepancy between the two periods leads to a non-closed rotational trajectory of two vortices in the $xy$ plane.
The most intriguing aspect is that the shape of the trajectory qualitatively changes depending on the initial position of the quantized vortices. 
When the vortices are initially positioned near the center, they rotate along an elliptical orbit; otherwise, they spiral outward with fluctuations. 
The critical distance at which the trajectory transitions from elliptical to spiral is comparable to the core size of the quantized vortices, suggesting that inter-vortex interactions play a role in this transition.
Notably, the vortex cores bend from a straight line along the rotational axis, but we infer that this deformation does not significantly contribute to the trajectory transition based on an analogy with the corotation in a quasi-two-dimensional dipolar BEC \cite{zhao2021, zhao2022}.

As the two quantized vortices corotate, the self-gravitating BEC deforms and rotates due to the superposition of multiple collective modes. 
The effective sizes of the BEC in each direction allow us to identify three distinct collective modes: the breathing mode, the quadrupole mode along the $z$-axis, and the quadrupole mode in the $xy$-plane.
Focusing on the periods of these collective modes, their dependencies on the initial position of the quantized vortices are different. 
The periods of the breathing mode and the quadrupole mode along the $z$-axis remain nearly constant, regardless of the initial vortex positions.
In contrast, the farther the quantized vortices are initially located from the center, the longer the period of the quadrupole mode becomes, gradually approaching the period of the quadrupole mode observed in the absence of quantized vortices.
The amplitudes of these collective modes are also influenced by the initial positions of the quantized vortices. 
Particularly, the quadrupole mode in the $xy$ plane becomes progressively amplified over time when the corotation of the two quantized vortices results in outward spiral trajectories.

We conclude that the effective dissipation of the quantized vortices is triggered by the coupling between the radial fluctuations of the vortices and the quadrupole mode, resulting from the transfer of energy from the vortices to the BEC through excited phonons.
When the two quantized vortices are initially positioned sufficiently far from the center, their radial oscillations resonate with the quadrupole mode in the $xy$ plane due to the coincidence of their periods.
The kinetic energy components reveal that the quantized vortices experience energetic damping, leading to phonon excitation within the BEC.
Consequently, the two rotating quantized vortices spiral out and the quadrupole mode is amplified.
This energy transfer does not occur when the two quantized vortices are initially positioned either near the center or near the surface of the BEC.
In the former case, this is due to the disparity between the periods of the two oscillations, while in the latter case, it is likely because the corotation of the two quantized vortices is strongly coupled with the quadrupole mode.

Based on the findings of this work, we can find that a self-gravitating BEC returns energy to the two corotating quantized vortices after initially receiving energy from them. 
As shown in Fig. \ref{figure4} (e) and (f), when the initial positions of the two quantized vortices are sufficiently separated, their corotation transitions to an inward spiral after initially spiraling outward to a certain degree.
This switching of trajectory indicates a reversal in the direction of energy transfer, with the quantized vortices receiving energy from the BEC. 
Notably, since the total angular momentum is conserved within this system, we can anticipate that the energy transfer periodically alternates.

Such corotation of two quantized vortices in a self-gravitating BEC can occur following the merger of multiple BECs.
When two self-gravitating BECs collide, the resulting BEC often exhibits elliptical deformation accompanied by rotation \cite{schwabe2016}. 
Furthermore, as observed in this paper, a cross-sectional view of the elliptical density profile reveals the presence of a few vortex-like structures.
These features align with our numerical findings in this study.
Therefore, we infer that a self-gravitating BEC after merging is likely to exhibit deformation and rotation driven by collective modes, accompanied by the corotation of two quantized vortices.
This post-merger rotation of the self-gravitating BEC and the corotation of the quantized vortices within it could be connected to turbulence within a self-gravitating BEC \cite{mocz2018, liu2023} and may be related to the characteristic rotation of galaxies implying the existence of dark matter \cite{Binney, Rubin1970}.

This work leaves some open questions for the corotation of quantized vortices in a self-gravitating BEC.
First, while we change only the initial position of the quantized vortices, the dependencies on the strength of the gravitational interaction have yet to be revealed. 
The strength of the gravitational potential should influence the rotational period of quantized vortices in a self-gravitating BEC by analogy with that in a dipolar BEC.
Our result suggests that phonons mediate the energy exchange between the radial oscillations of the quantized vortices and the quadrupole oscillations of the self-gravitating BEC.
Thus, similar to the excitation of Kelvin waves in trapped atomic BECs \cite{mizushima2003}, a certain type of phonon decay (or synthesis) may occur.

\begin{acknowledgments}
This work was supported by JST, the establishment of university fellowships towards the creation of science technology innovation, Grant Number JPMJFS2138.
M. T. acknowledges the support from JSPS KAKENHI Grant Number JP23K03305 and JP22H05139.
\end{acknowledgments}

\appendix

\section{ANALYTICAL CALCULATION OF TWO PARALLEL QUANTIZED VORTICES IN A CIRCULAR CONTAINER BY VORTEX POINT MODEL}
We employ a vortex-point model with friction as a simplified approach to examine the dynamics of two corotating quantized vortices. 
This model is derived from the vortex filament model, commonly used in the study of quantized vortices in superfluid $^4$He \cite{Donnelly}.
Assuming that the vortex core is sufficiently small, the quantized vortex can be regarded as a filament parametrized by $\xi$.
The vortex filament moves within the superfluid flow field  $\bm{v}_s(\bm{r},t)$, and as a result it obeys the equation of motion 
\be
\f{d \bm{s}}{dt}
=
\bm{v}_s
-
\alpha\bm{s}'\times\bm{v}_s
+
\alpha'
\bm{s}'
\times
\left(
\bm{s}'\times\bm{v}_s
\right),
\label{schwartz}
\ee
using the position of the filament $\bm{s}(\xi,t)$ with the one-dimensional coordinate $\xi$ and $\bm{s}'(\xi,t)\equiv\f{\p\bm{s}(\xi,t)}{\p\xi}$.
The parameters $\alpha$ and $\alpha'$ imply the strength of dissipation and in the case of superfluid $^4$He, they give rise to phenomenological friction between the superfluid and normal-fluid components, so-called "mutual friction".
However, we assume these parameters do not vanish in Eq. (\ref{schwartz}) despite no normal fluid component in order to compare the result to our numerical findings on a self-gravitating BEC.
In other words, these dissipations are caused by different mechanisms from the superfluid $^4$He.
For convenience, we neglect $\alpha'$ below that.

In an incompressible quantum fluid, the vortex filament $L$ induces the superfluid velocity field $\bm{v}_s(\bm{r},t)$ following the Biot-Savart law written by
\be
\bm{v}_s(\bm{r},t)
=
\f{\kappa}{4\pi}
\int_L
d\xi
\f{\bm{s}'(\xi,t)\times(\bm{r}-\bm{s}(\xi,t))}{\left|\bm{r}-\bm{s}(\xi,t)\right|^3}.
\label{bs_law}
\ee
Specifically, a straight vortex line with circulation $\kappa$ generates a counter-clockwise velocity field given by 
\be
v_s
=
\f{\kappa}{2\pi r}
\label{velocity_field}
\ee
around the vortex line, where $r$ denotes the distance from the vortex core. 
Assuming the vortex lines are straight in a cylindrically symmetric system, the system can be reduced to two dimensions, where the vortex lines are treated as points.

Considering quantized vortices are in a container, the boundary effect can be replaced by introducing an image vortex positioned at the corresponding location. 
In this work, we specifically treat the scenario in which the quantized vortices are confined within a circular container of radius $R$.
The circular container imposes boundary conditions such that the radial component of the velocity field vanishes. 
Hence, when a vortex point is situated at a distance $r_v$ from the center of the container, the image vortex, carrying a charge of $-\kappa$, is positioned in the same direction as the quantized vortex, with a distance from the center of the container given by $R^2/r_v$.
 
We consider that two vortex points with the same charges are symmetrically located in the container.
While one of them is positioned far from the center of the container with a distance $r_v$ and an angle $\theta$, the other is positioned with the common distance in the opposite direction.
Then, a vortex point moves in the flow field generated by the other and two image vortex points.
Using Eqs. (\ref{schwartz}) and (\ref{velocity_field}), the equation of motion is written by
\begin{subequations}
\begin{eqnarray}
\f{dr_v}{dt}=\f{\alpha\kappa}{2\pi}
\left\{
\f{1}{2r_v}
+
\f{r_v}{R^2-r_v^2}
-
\f{r_v}{R^2+r_v^2}
\right\}, 
\label{appa}
\\
r_v\f{d\theta}{dt}=\f{\kappa}{2\pi}
\left\{
\f{1}{2r_v}
+
\f{r_v}{R^2-r_v^2}
-
\f{r_v}{R^2+r_v^2}
\right\}.
\label{appb}
\end{eqnarray}
\end{subequations}

\begin{figure}
\centering
\includegraphics[width=8.5cm]{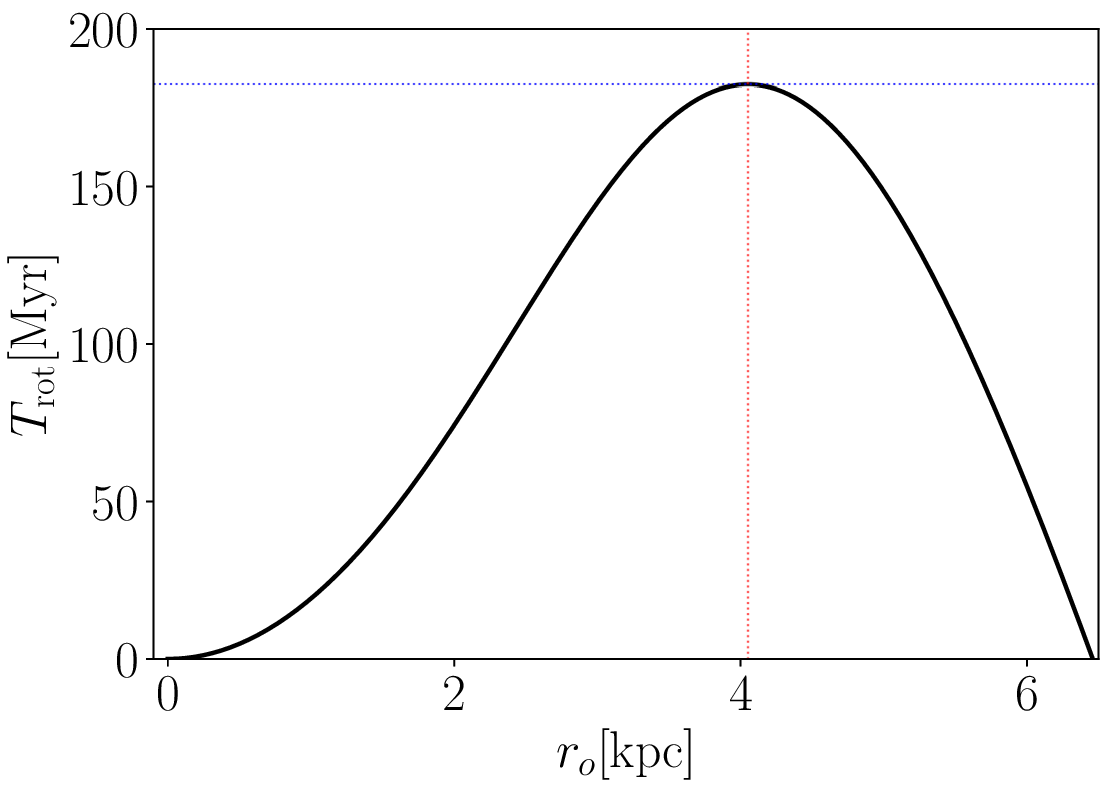} 
\caption{
The rotational period of the two quantized vortices corotating in a circular container in Eq. (\ref{period_sub}) when $R=6.46$ kpc and $\kappa=4.10~\textrm{kpc}^2/\textrm{Myr}$.
The red dotted line shows the radius where the period becomes maximum and the blue dotted line indicates this value.
}
\label{figureA1}
\end{figure}
Without dissipation of the quantized vortices, namely $\alpha=0$, we find that the vortex points do not displace radially from the initial positions, as $\f{dr_v}{dt}=0$, derived from Eq. (\ref{appa}).
Equation (\ref{appb}) then suggests that the vortex points rotate with a constant period
\begin{equation}
T_\mathrm{rot}
=
\f{8\pi^2r_o^2}{\kappa}
\f{R^4-r_o^4}{R^4+3r_o^4}
\label{period_sub}
\end{equation}
around the center of the container, where the initial distance of $r_v$ is represented as $r_o$.
Consequently, the two vortex points undergo uniform circular motion along the same orbit.
The $r_o$-dependence of this rotational period is shown in Fig. \ref{figureA1}.
In this figure, $\kappa=4.10~\textrm{kpc}^2/\textrm{Myr}$ is set to the same value as that used for the self-gravitating BEC in this work and $R=6.46$ kpc corresponds to the Thomas-Fermi radius of this BEC to facilitate comparison to our numerical result.
When the vortex points are near the center of the container, the period increases quadratically with $r_o$, similar to that in the uniform case given by
\be
T_\mathrm{rot}=\f{8\pi^2r_o^2}{\kappa}
\label{period_uniform}
\ee
with $R\rightarrow\infty$ in Eq. (\ref{period_sub}).
However, as $r_o$ further increases, the period no longer shows sufficient increase and gradually decreases near the boundary due to the influence of the image vortices, \textit{i.e.}, the finite-size effects.
As a reslt, the period reaches a maximum value $T_\mathrm{rot}=\sqrt{\f{2-\sqrt{3}}{3\sqrt{3}}}\f{8\pi^2}{\kappa}R^2\approx182~\textrm{Myr}$ when $r_o=\{\f{2}{\sqrt{3}}-1\}^{\f{1}{4}}R\approx4.05~\textrm{kpc}$, as indicated by the red and blue dotted lines.

When the dissipation of the quantized vortices is present, namely $\alpha >0$, the two vortex points move outward while rotating within the container.
Solving Eq. (\ref{appa}), $r_v$ satisfies
\be
\begin{split}
\f{\alpha\kappa}{2\pi} t
=
&\f{4\sqrt{3}}{9}R^2
\tan^{-1}\left(\f{\sqrt{3}r_{v}^2}{R^2}\right) \\
&-
\f{4\sqrt{3}}{9}R^2
\tan^{-1}\left(\f{\sqrt{3}r_{o}^2}{R^2}\right) \\
&-
\f{r_{v}^2-r_o^2}{3},
\label{Y_time}
\end{split}
\ee
which demonstrates that $r_v$ monotonically increases with time until the vortex points reach the boundary of the container.
Furthermore, Eq. (\ref{appb}) indicates that the angular velocity also varies over time, initially decreasing and accelerating as the vortex points approach the boundary.

\end{document}